\documentclass[%
 reprint,
 amsmath,amssymb,
 aps,
 prl
]{revtex4-2}

\usepackage[caption=false]{subfig}
\usepackage{pythonhighlight}

\usepackage {xcolor}

\usepackage[hidelinks]{hyperref}
\usepackage{graphicx}
\usepackage{dcolumn}
\usepackage{bm}

\DeclareMathOperator{\Tr}{Tr}

\begin{document}

\preprint{APS/123-QED}

\title{Speed limit,  dissipation bound and dissipation-time trade-off in thermal relaxation processes}

\author{Jie Gu}
 \affiliation{Chengdu Academy of Education Sciences, Chengdu 610036, China}
\email{jiegu1989@gmail.com}
\date{\today}

\begin{abstract}
We investigate bounds on speed, non-adiabatic entropy production and trade-off relation between them for classical stochastic processes with time-independent transition rates. 
Our results show that the time required to evolve from an initial to a desired target state is bounded from below by the informational-theoretic $\infty$-R\'enyi divergence between these states, divided by the total rate. 
Furthermore, we conjecture and provide extensive numerical evidence for an  information-theoretical bound on the non-adiabatic entropy production and a novel dissipation-time trade-off relation that outperforms previous bounds in some cases.
\end{abstract}

\maketitle


\emph{Introduction.---}
Optimal control of a system’s evolution from an initial to a desired target state is a crucial task  \cite{Glaser2015,Deffner2020,Koch2022,blaber2023}, with close ties to the optimal transport problem \cite{villani2009optimal,vanvu2023}. The definition of ``optimal” varies depending on the specific cost function employed, which may include time, energy consumption, dissipation, error, robustness, or trade-offs between them. 
In addition to optimal control protocols, non-model-specific fundamental bounds on the cost functions are of significant interest \cite{gong2022}. 
For instance, in quantum systems, rapid state transformations are usually desirable, thereby motivating extensive investigations of the so-called ``quantum speed limit" \cite{Mandelstam1945,Margolus1998,Deffner2017}.
For a quantum system with a time-independent Hamiltonian $H$, the time it needs to evolve from the initial state $\rho^\text{(i)}$ to the final state $\rho^\text{(f)}$ is bounded from below by ($\hbar =1$)
\begin{equation}
\label{eq:qsl}
\tau \ge \max \left \{  \frac{\mathcal{L}(\rho^\text{(i)}, \rho^\text{(f)})}{\Delta H}, \frac{2\mathcal{L}^2(\rho^\text{(i)}, \rho^\text{(f)})}{\pi \langle H \rangle}  \right \},
\end{equation}
where $\mathcal{L}(\rho^\text{(i)}, \rho^\text{(f)})$ is the Bures angle that quantifies the distance between the endpoints, and $\Delta H$ and $\langle H \rangle$ are variance and average of the Hamiltonian, respectively \cite{giovannetti2003,giovannetti2004}.

Recent developments have extended the concept of speed limits to classical stochastic processes, where entropy production  plays a crucial role \cite{shanahan2018, Okuyama2018,Shiraishi2018, Vo2020,Lee2022}\footnote{
There exist alternative speed limits that are not explicitly defined in terms of entropy production. For instance,  the relevant quantity could be the rate of change in the information content \cite{nicholson2020,garcia-pintos2022}, the entropy flux \cite{falasco2020}, or the dynamical activity \cite{dechant2022a}. 
}.  For Markovian stochastic processes with given initial and final probability distributions $\boldsymbol{p}^\text{(i)}$ and $\boldsymbol{p}^\text{(f)}$, speed limits can be expressed as a trade-off between entropy production $\Sigma$ and time duration $\tau$ given by ($k_\text{B}=1$)
\begin{equation}
\label{eq:general}
    \Sigma \ge {f_{\boldsymbol{p}^\text{(i)} \to \boldsymbol{p}^\text{(f)}}(\mathcal{R} \tau) },
\end{equation}
where $f$ is a monotonically decreasing function of $\mathcal{R}\tau$. 
The subscript $\boldsymbol{p}^\text{(i)} \to \boldsymbol{p}^\text{(f)}$ denotes the dependence of the function on the endpoints, and $\mathcal{R}$ quantifies the system's timescale.  This inequality encompasses the following three important cases in the literature.

In generic processes entropy production  can be split into adiabatic  and non-adiabatic (Hatano-Sasa \cite{Hatano2001}) contributions \cite{esposito2010}. The former persists even when the probability distribution equals the instantaneous steady state, while the latter arises from deviations from this state. 
It was reported that non-adiabatic  entropy production $\Sigma_\text{na}$ is lower-bounded by (referred to as activity bound) \cite{Shiraishi2018, Vo2020}
\begin{equation}
\label{eq:DA}
\Sigma\ge \Sigma_\text{na} \ge \frac{\mathcal{L}^2(\boldsymbol{p}^\text{(i)}, \boldsymbol{p}^\text{(f)})}{2\mathcal{A} \tau}   ,
\end{equation}
where $\mathcal{L}(\boldsymbol{p}^\text{(i)}, \boldsymbol{p}^\text{(f)})$ represents the total-variation distance between endpoints and $\mathcal A$ denotes time-averaged dynamical activity \cite{Maes2020}. 

In the quasi-static limit with $\tau \to \infty$, the time-averaged dynamical activity  becomes time-independent, resulting in an inequality $\Sigma_\text{na} \ge \mathcal{O}(1/\tau)$.
This observation bears resemblance to the result for slow but finite-time Markovian stochastic processes \cite{Sinitsyn2007,Sinitsyn2009,Ren2010,sinitsyn2011,Gu2017}, where the entropy production bound is given by $  \Sigma \ge {\mathcal{L}_\text{T}^2}/{\tau}
$. Here $\mathcal{L}_\text{T}$ denotes the thermodynamic length \cite{Kirkwood1946,Weinhold1975,Ruppeiner1979,Salamon1983,Janyszek1989,Brody1995,Ruppeiner1995,Sivak2012,Zulkowski2012,machta2015, Mandal2016,Miller2019,Miller2020,Scandi2019,Scandi2020, Abiuso2020,Frim2022,gu2023}, which already encodes the information about the time scale derived from the transition rates.

For relaxation processes with a time-independent transition rate matrix satisfying the detailed balance condition,   the general dissipation-time trade-off relation Eq. \eqref{eq:general}  is superseded by a simpler $\tau$-independent bound  \cite{Shiraishi2019}
\begin{equation}
\label{eq:divergencebound}
\Sigma \ge  D_1(\boldsymbol{p}^\text{(i)}\|\boldsymbol{p}^\text{(f)}),    
\end{equation}
where \(D_1(\boldsymbol{p}^\text{(i)}\|\boldsymbol{p}^\text{(f)})\) is the $1$-R\'enyi divergence \cite{VanErven2014,sagawa2022entropy},  also known as  Kullback-Leibler divergence or relative entropy \cite{Vedral2002}.
Since the bound is independent of time duration, the trade-off  between dissipation and evolution time is, in fact, concealed.
In the context of this kind of simple but important processes, i.e., relaxation processes with time-independent transition rates where the detailed balance condition is not necessarily met \cite{van1992}, three natural questions arise.
First, given the similarity between a quantum process with a time-independent Hamiltonian and a thermodynamic process with a time-independent transition matrix, one may wonder whether there exists a speed limit analogous to Eq. \eqref{eq:qsl}, where the bound is given by the distance between the endpoints divided by a timescale constant \footnote{There exists a relevant speed limit for  open quantum systems governed by a
Markovian quantum master equation \cite{delcampo2013}. Surprisingly,  the bound therein only involves the initial state and the dynamical map.}. 
Second, it is tempting to see whether Eq. \eqref{eq:divergencebound} also holds for the non-adiabatic entropy production.
Third, one may inquire the trade-off relation in the form of Eq. \eqref{eq:general}, if any, between dissipation and time in these relaxation processes.
Specifically, it is desirable to obtain a suitable timescale constant.

In this Letter, we answer these three questions  and show that the relevant timescale constant  is the total rate, i.e., the sum of all positive transition rates. We also demonstrate with examples that our new dissipation-time trade-off relation outperforms previous bounds.

\emph{Speed limit.---}
Consider a stochastic   Markov jump process with finite $N$ states.
The dynamics of the probability distribution
\({\boldsymbol{p}}=[p_1,p_2,\ldots, p_N]^\text{T}\) is described by a Pauli master equation \cite{van1992}
\begin{equation}
\label{eq:me}
    \frac{\text{d} {p}_{m}}{\text{d} t}=\sum_{n} W_{mn} p_{n},
\end{equation}
where \(p_m\) is the probability of state \(m\), \(W_{mn} (m\ne n)\) is the
time-{independent} transition rate from state \(n\) to \(m\), and
\(W_{mm} = -\sum_{n,n\ne m} W_{nm}\).
For later use, we define the \emph{total rate} as
\begin{equation}
    \mathcal{W} = \sum_{m \ne n} W_{mn} = -\mathrm{Tr} \mathbf{W}.
\end{equation}
In experiments,  the transition rate matrix and its trace can be inferred from trajectory data \cite{bladt2005}.
Provided that the Markov chain is ergodic, a steady state distribution $\boldsymbol{p}^\text{(ss)}$ is expected, satisfying $\sum_n W_{mn}p_n^\text{(ss)}=0$. Steady states can be divided into two categories depending on whether they meet the detailed balance condition, $W_{mn}p_n^\text{(ss)} = W_{nm}p_m^\text{(ss)}$, which is not assumed throughout this Letter.
The total entropy production $\Sigma$ consists of an adiabatic contribution and a non-adiabatic one, both of which are non-negative.
Given an
initial state \(\boldsymbol{p}^\text{(i)}\) and a target state
\(\boldsymbol{p}^\text{(f)}\), the non-adiabatic entropy production is given by \cite{broeck2013,Shiraishi2018}
\begin{equation}
\label{eq:na}
    \Sigma_\text{na}  = D_1(\boldsymbol{p}^\text{(i)} \| \boldsymbol{p}^\text{(ss)})-D_1(\boldsymbol{p}^\text{(f)} \| \boldsymbol{p}^\text{(ss)}) ,
\end{equation}
where \({D}_1(\boldsymbol{p} \|\boldsymbol{q}) = \sum_n p_n \ln(p_n/q_n)\) is the $1$-R\'enyi divergence 
between the two probability distributions \cite{VanErven2014,sagawa2022entropy,Vedral2002}.
For arbitrary dynamics, $\Sigma \ge \Sigma_\text{na}$, and the equality is attained when the detailed balance condition is satisfied.
Notably,  the detailed balance condition is always fulfilled in two-state systems.

The fixed endpoints impose constraints on the transition rate matrix $\boldsymbol{W}$, but generally  do not uniquely determine it.
The choice of $\boldsymbol{W}$ affects the magnitude of entropy $\Sigma$ and $\Sigma_\text{na}$ produced by the stochastic process connecting the same endpoints over time $\tau$.
For two-state systems, the steady state  $\boldsymbol{p}^\text{(ss)}$ is uniquely determined, with $p_1^\text{(ss)}$ expressible in terms of $\mathcal{W} \tau$ as \footnote{See the Supplemental Material for derivation of Eq. \eqref{eq:p1}  and the code to verify Eqs.  \eqref{eq:boundEPna} and \eqref{eq:newbound}.}
\begin{equation}
\label{eq:p1}
   p_1^\text{(ss)} = p_1^\text{(i)} + \frac{p_1^\text{(f)}-p_1^\text{(i)}}{1-e^{-\mathcal{W} \tau}} ,
\end{equation}
with $p_2^\text{(ss)}=1-p_1^\text{(ss)}$.

\begin{figure*}[!t]
\centering
     \subfloat[]{\includegraphics[width=0.33\textwidth]{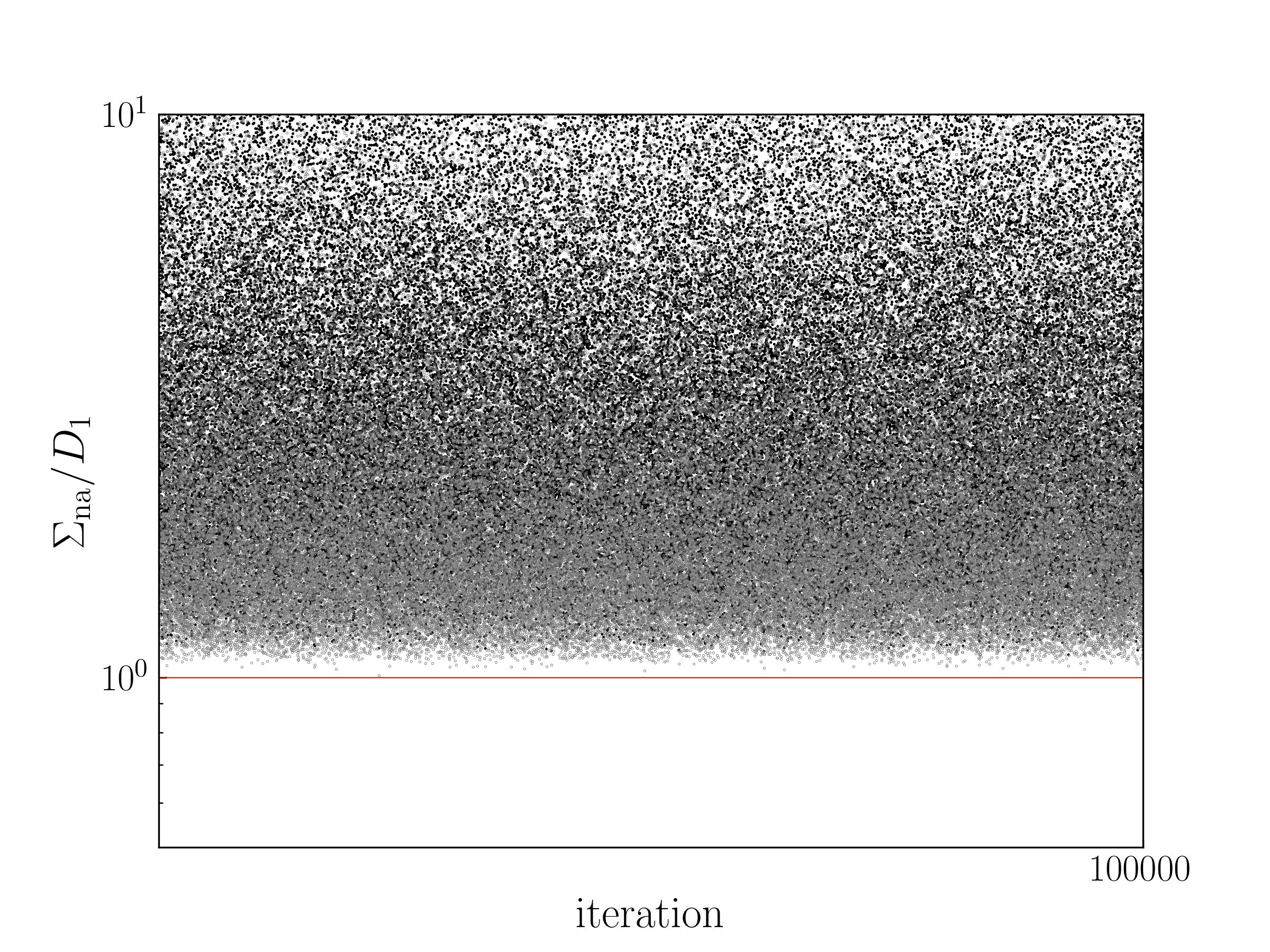} } 
     \subfloat[]{\includegraphics[width=0.33\textwidth]{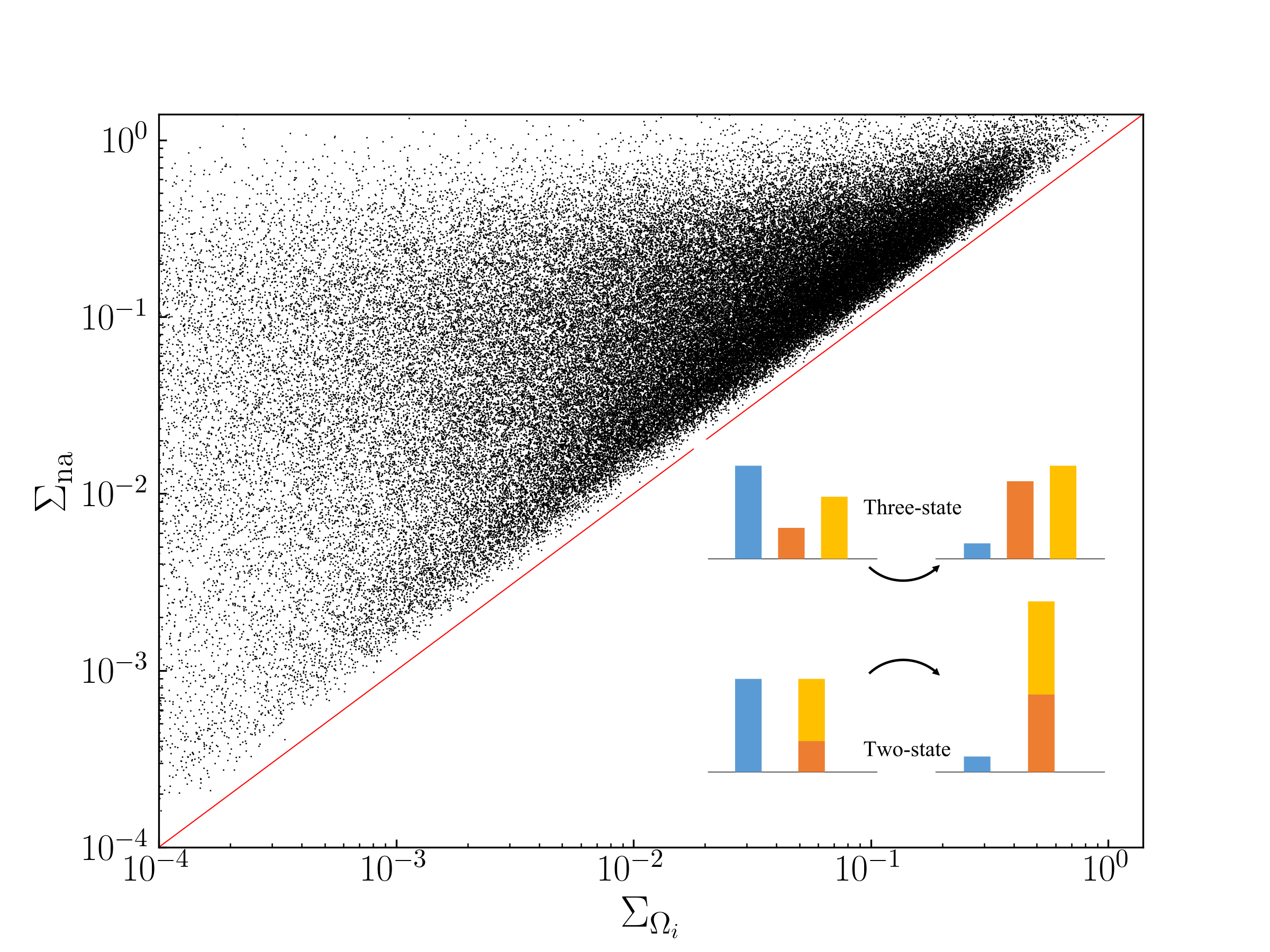} }  
     \subfloat[]{\includegraphics[width=0.33\textwidth]{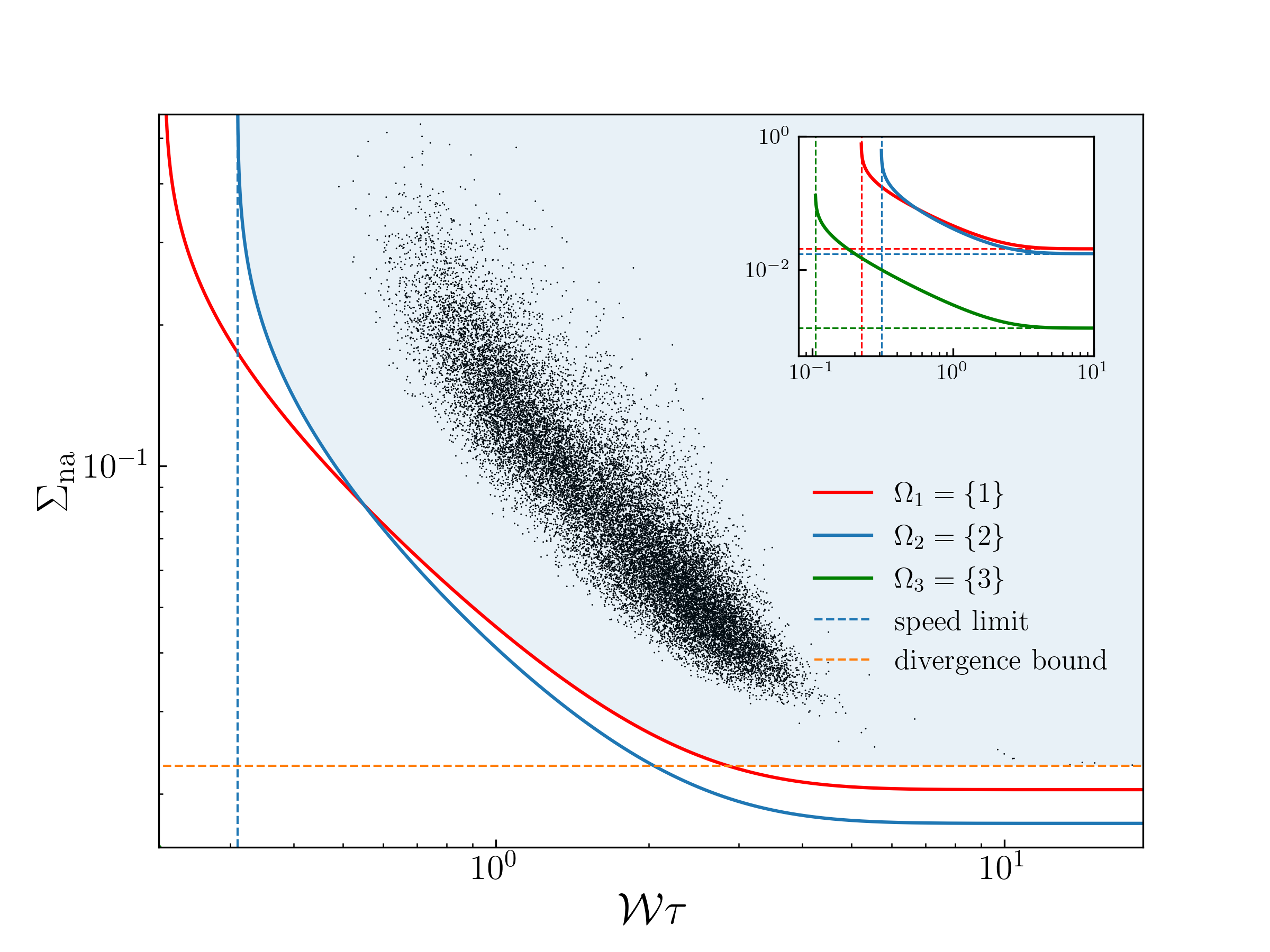} }
    \caption{
\label{fig1}
(a) $\Sigma_\text{na}(\tau)/D_1(\boldsymbol{p}^\text{(i)},\boldsymbol{p}^\text{(f)})$ represented by black dots and $\Sigma_\text{na}(\tau/2)/D_1(\boldsymbol{p}^\text{(i)},\boldsymbol{p}^\text{(f)})$ represented by gray circles.
(b) Non-adiabatic entropy production $\Sigma_\text{na}$ vs. corresponding two-state entropy production $\Sigma_{\Omega_i}$.
The inset depicts a schematic of the pseudo-coarse-graining procedure.
(c) Non-adiabatic entropy production $\Sigma_\text{na}$ vs. dimensionless time duration $\mathcal{W}\tau$, with the endpoints fixed.
The inset presents the two-state curves for different partitioning schemes, and the horizontal and vertical dashed lines represent the asymptotes.
In both (a) and (b),  there are $10^5$ data points, with each generated as follows. An initial three-state probability distribution is sampled randomly, together with a transition rate matrix with positive entries drawn uniformly from $[0,1]$. An evolution time $\tau$ is also drawn uniformly from $[0,1]$, then the state at $t=\tau/2$ and the final state are calculated. Using Eq. \eqref{eq:na}, we compute the non-adiabatic entropy production $\Sigma_\text{na}(\tau/2)$ and $\Sigma_\text{na}(\tau)$.  
We then randomly partition the states into two disjoint nonempty sets and use Eqs. \eqref{eq:p1} and \eqref{eq:na} to calculate $\Sigma_{\Omega_i}$.
In (c) there are 40000 data points,  each corresponding to a transition from $\boldsymbol{p}^\text{(i)} = [0.5,0.3,0.2]^\text{T}$ to $\boldsymbol{p}^\text{(f)} = [0.6,0.22,0.18]^\text{T}$ with different $\boldsymbol{W}$ and $\tau$ obtained as follows: We randomly select a transition rate matrix $\boldsymbol{W}$ and an evolution time $\tau$ as initial guesses and minimize the total-variation distance between $\exp(\boldsymbol{W} \tau) \boldsymbol{p}^\text{(i)}$ and $\boldsymbol{p}^\text{(f)}$ with the error threshold set to $10^{-6}$. The optimal values of $\boldsymbol{W}$ and $\tau$ are then used to calculate and plot the non-adiabatic entropy production as a function of $\mathcal{W\tau}$.
    }
\end{figure*}

By substituting Eq. \eqref{eq:p1} into  \eqref{eq:na}, it can be observed that, given the endpoints, the non-adiabatic entropy production in a two-state system is a monotonically decreasing function of $\mathcal{W}\tau$. 
See the inset of Fig. \ref{fig1} (c).
In other words, the $\Sigma$ versus $\mathcal{W}\tau$ curve indicates a trade-off between entropy production and  time duration, and is bounded by a vertical and a horizontal asymptote.
The horizontal asymptote signifies that the minimum of the entropy production is reached when $\mathcal{W}\tau \to \infty$. 
In this limit, $\boldsymbol{p}^\text{(ss)} = \boldsymbol{p}^\text{(f)}$, so Eq. \eqref{eq:na} implies that the minimum entropy production is given by \eqref{eq:divergencebound}.


It is unphysical for the population $p_1^\text{ss}$ to be $0$ or $1$, as this would require one of the transition rates to vanish.
Combining Eq. \eqref{eq:p1} with this constraint ($0<p^\text{(ss)}_1<1$), we obtain a speed limit represented by the vertical asymptote, i.e., a lower bound for the evolution time $\tau$:
\begin{equation}
\label{eq:speedlimit2}
    \tau >  \frac{ \max \{ \ln (p_1^\text{(i)}/p_1^\text{(f)} ), \ln (p_2^\text{(i)}/p_2^\text{(f)} )\} }{\mathcal{W}}.
\end{equation}
The numerator coincides with the $\infty$-R\'enyi divergence, defined as $D_\infty(\boldsymbol{p} \| \boldsymbol{q})=\max_n\left\{ \ln {p_n}/{q_n} \right\}$ \cite{VanErven2014,sagawa2022entropy}.
Specifically, the lower bound on evolution time is given by the quotient of the $\infty$-R\'enyi divergence between the two endpoints and the total rate of transitions $\mathcal{W}$.

This lower bound reflects an information-theoretical limit on speed in two-state systems; hence, it is relevant to examine whether this bound can be extended to generic $N$-state systems.
The answer is affirmative. 
We state our main result here while deferring the proof to the end of the Letter: the time it needs to evolve from the initial state $\boldsymbol{p}^\text{(i)}$ to the final state $\boldsymbol{p}^\text{(f)}$ is bounded from below by
\begin{equation}
\label{eq:speedlimit}
    \tau >  \frac{ D_\infty(\boldsymbol{p}^\text{(i)} \| \boldsymbol{p}^\text{(f)}) }{\mathcal{W}},
\end{equation}
where $D_\infty(\boldsymbol{p}^\text{(i)} \| \boldsymbol{p}^\text{(f)})$ is the $\infty$-R\'enyi divergence between the distributions $\boldsymbol{p}^\text{(i)}$ and $\boldsymbol{p}^\text{(f)}$, and $\mathcal{W}$ is the total rate. Remarkably, this lower bound on evolution time resembles the distance-based formulation in Eq. \eqref{eq:qsl}, where a constant rate characterizes the timescale. It is noteworthy that the 
$\infty$-R\'enyi divergence plays a significant role in the resource theory of thermodynamics \cite{horodecki2013,aberg2013,sagawa2022entropy}, and its application to speed limits highlights its versatility and significance in various physical contexts.
If $\boldsymbol{p}^\text{(f)} = \boldsymbol{p}^\text{(ss)} $, the time it needs to reach the steady state is infinite, so this bound is trivially satisfied.
When the final state $\boldsymbol{p}^\text{(f)}$ is not of full rank, both the $\infty$-R\'enyi divergence and the bound of time diverge. This is a manifestation of the third law of thermodynamics, which states that non-full rank states cannot be attained in finite time \cite{masanes2017}.

\emph{Information-theoretical bound on non-adiabatic entropy production.---}
Building upon the information-theoretical bound on entropy production described in Ref. \cite{Shiraishi2019} for systems with detailed balance, we extend our findings to scenarios without detailed balance. Our second result introduces a conjecture (referred to as divergence bound):
\begin{equation}
\label{eq:boundEPna}
\Sigma_\text{na} \ge D_1(\boldsymbol{p}^\text{(i)}\|\boldsymbol{p}^\text{(f)}).
\end{equation}
This conjecture, presented as an extension of Eq. \eqref{eq:divergencebound}, suggests a lower bound on the non-adiabatic entropy production given by the $1$-R\'enyi divergence between the endpoints.
Fig. \ref{fig1}(a) provides numerical evidence for three-state systems to support it, and the code used to generate numerical verification for other numbers of states can be found in the Supplemental Material \cite{Note3}.
In fact we confirm a tighter bound, $\Sigma_\text{na}(\tau/2) \ge D_1(\boldsymbol{p}^\text{(i)}\|\boldsymbol{p}^\text{(f)})$, where $\Sigma_\text{na}(\tau/2)$ represents the non-adiabatic entropy production during the time interval $[0,\tau/2]$, which strengthens our conjecture.
Our findings have implications for the excess entropy production $\Sigma_\text{ex}$,  which is defined by a variational principle and is always greater than $\Sigma_\text{na}$ \cite{kolchinsky2022,shiraishi2023a}. 
As a consequence, we observe the following relationships:
\begin{equation}
\Sigma_\text{ex}(\tau) \ge \Sigma_\text{ex}(\tau/2) \ge D_1(\boldsymbol{p}^\text{(i)}\|\boldsymbol{p}^\text{(f)}),
\end{equation}
as reported in Refs. \cite{kolchinsky2022,shiraishi2023a}.

\emph{Trade-off between dissipation and time.---} 
Let us state  the third main result: for $N$-state dynamics evolving from $\boldsymbol{p}^\text{(i)}$ to $\boldsymbol{p}^\text{(f)}$ during $\tau$ with an total rate $\mathcal{W}$, the conjecture is that the entropy production $\Sigma$ is bounded from below by (referred to as trade-off bound)
\begin{equation}
\label{eq:newbound}
\Sigma_\text{na} \ge \Sigma_{\Omega_i}(\mathcal W \tau) \quad i =1,2,\ldots,S_N^{(2)}.
\end{equation}
Here , $\Sigma_{\Omega_i}(\mathcal W \tau)$ is the entropy production of two-state dynamics obtained from a pseudo-coarse-graining procedure as follows:  partition the set of $N$ states into two nonempty sets where the first set is denoted by $\Omega_i$, 
then we have the two endpoints
$\boldsymbol{P}_{\Omega_i}^\text{(i)} = [\sum_{n\in {\Omega_i}}p_n^\text{(i)},1-\sum_{n\in {\Omega_i}}p_n^\text{(i)}]^\text{T}$ to $\boldsymbol{P}_{\Omega_i}^\text{(f)} = [\sum_{n\in {\Omega_i}}p_n^\text{(f)},1-\sum_{n\in {\Omega_i}}p_n^\text{(f)}]^\text{T}$.
We then use Eq. \eqref{eq:p1} to find the steady state distribution, where $\mathcal{W}$ and $\tau$ of the original $N$-state dynamics are used.
The entropy production of the two-state dynamics is obtained by substituting the steady state distribution into  \eqref{eq:na}.
The inset of Fig. 1(b) gives a schematic of the pseudo-coarse-graining procedure for three states. 
As shown in the upper panel, the  system evolves from $\boldsymbol{p}^\text{(i)} = [0.5,0.17,0.33]^\text{T}$ to $\boldsymbol{p}^\text{(f)} = [0.1,0.4,0.5]^\text{T}$, visually depicted by bars of varying colors and lengths. 
There  are three partitioning schemes and the lower panel shows one in which the second and third state are grouped together, resulting in
 $\boldsymbol{P}_{\Omega_1}^\text{(i)} = [0.5, 0.5]^\text{T}$ and $\boldsymbol{P}_{\Omega_1}^\text{(i)} = [0.1, 0.9]^\text{T}$.
Given $N$ states, we have $S_N^{(2)}$ different ways of partitioning the $N$ states into two nonempty sets, where $S_N^{(2)}$ is the Stirling number of the second kind \cite{Abramowitz1965}.
Explicitly,  $S_N^{(2)} =2^{N-1}-1$. For example,  four states can be partitioned in $S_4^{(2)}=7$ ways, where $\Omega_i=\{1\},\{2\},\{3\},\{4\},\{1,2\},\{1,3\},\{1,4\}$, with the elements in each set being the indices of states.
Thus, there are $S_N^{(2)}$ different two-state bounds in total, jointly bounding $\Sigma_\text{na}$.
As discussed above, these two-state bounds are monotonically decreasing function of $\mathcal{W} \tau$, which has the form of Eq. \eqref{eq:general}, with $\mathcal{W}$ playing the role of $\mathcal{R}$.
We stress that the two-state dynamics cannot be obtained from the standard coarse-graining procedure \cite{esposito2012,zhen2021}: the population ${P}_\Omega(t)$ is in general not equal to $\sum_{n \in \Omega} p_n(t)$ except at the two endpoints; the transition rate matrix for each two-state dynamics is time-independent, and its entries are not simple linear combinations of the original transition rates.
In the following, we consider the three-state case as an example, and numerical evidence to support Eq. \eqref{eq:newbound} for other numbers of states can be generated using the code in the Supplemental Material \cite{Note3}.

As a direct verification of the inequality \eqref{eq:newbound}, Fig. \ref{fig1}(b) shows a plot of the non-adiabatic entropy production $\Sigma_\text{na}$, versus the corresponding two-state $\Sigma_{\Omega_i}$. 
All data points lie above the diagonal, confirming the new bound given by \eqref{eq:newbound}.
In the long-$\tau$ limit, the entropy production of the two-state dynamics is $D_1(\boldsymbol{P}_{\Omega_i}^\text{(i)} \|\boldsymbol{P}_{\Omega_i}^\text{(f)} )$.
By applying the theorem that refinement cannot decrease divergence \cite{alajaji2018}, which is essentially the log-sum inequality, this quantity is not greater than ${D}_1(\boldsymbol{p}^\text{(i)} \|\boldsymbol{p}^\text{(f)})$ and also $\Sigma_\text{na}$. 
This leads to two questions: (1) Can the bound ever be saturated, as it is not evident from Fig. \ref{fig1}(a)? (2) Is the new bound ever tighter than the divergence bound and the activity bound, or is it otherwise redundant?

\begin{figure}[!ht]
\centering
\includegraphics[width=\columnwidth]{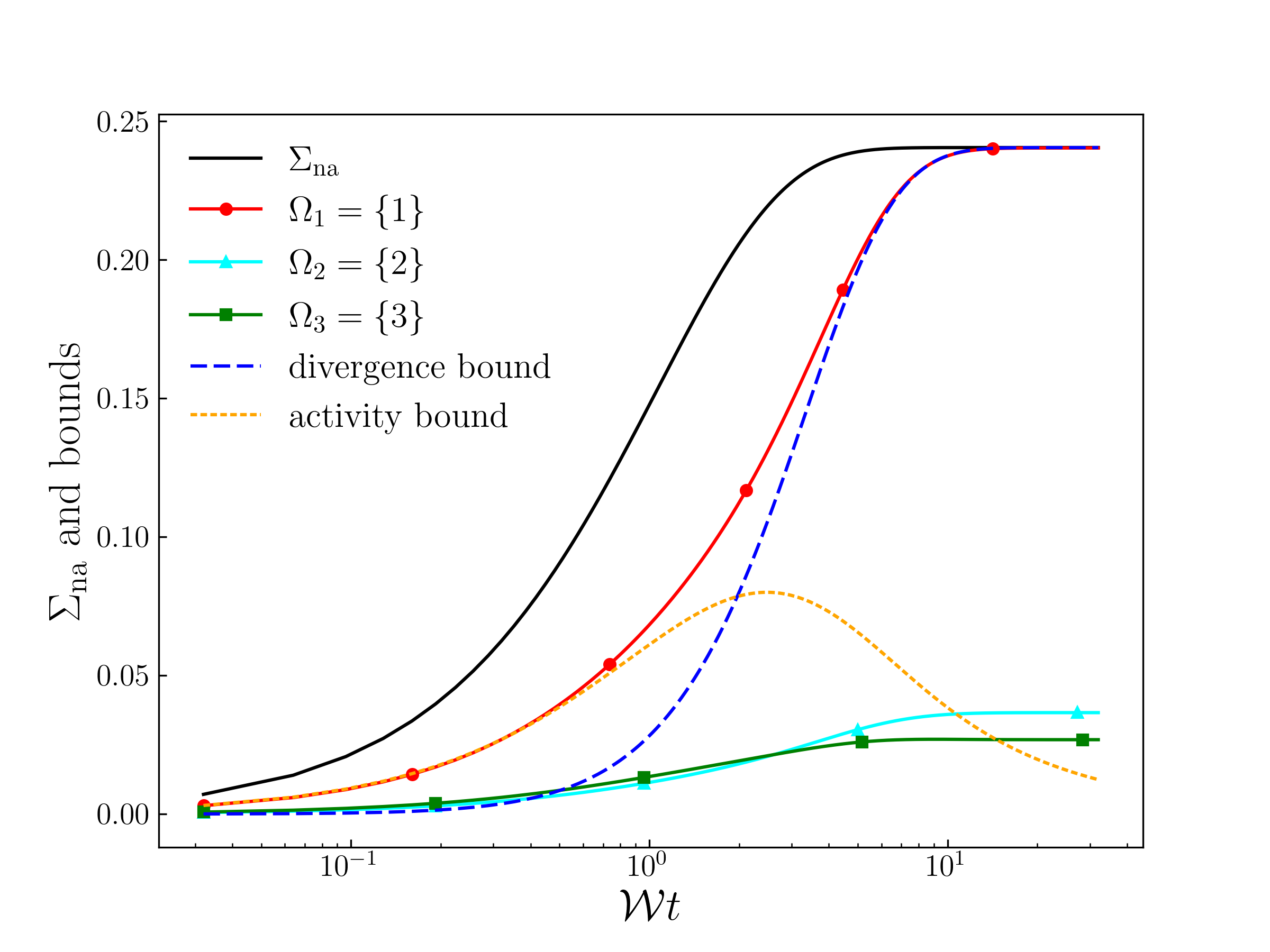}
\caption{
\label{fig3}
Non-adiabatic entropy production $\Sigma_\text{na}$, corresponding divergence bound \eqref{eq:divergencebound}, activity bound \eqref{eq:DA} and the trade-off bound \eqref{eq:newbound} as a function of the dimensionless time $\mathcal{W}t$. There are three ways to partition a set with three states: one of the two nonempty set is $\Omega_i=\{1\},\{2\}$ and $\{3\}$, respectively.
}
\end{figure}

There are at least two simple cases in which the bound is (nearly) saturated.
By examining the condition for equality in the log-sum inequality, it can be observed that as long as a subset $\Omega$ of the states such that $p^\text{(i)}_m/p^\text{(ss)}_m$ are equal for all $m \in \Omega$, the bound is saturated in the long-$\tau$ limit.
Another scenario in which the bound is nearly saturated throughout the entire process is when all the $N-2$ states' initial populations and the transition rates between them are vanishingly small, effectively reducing the dynamics to only two states.
This case also partially addresses the second question, as the divergence bound is not generally saturated.
The superiority of the new bound over the previous bounds is also evident in nontrivial cases, as will be seen in Figs. \ref{fig1}(c) and Fig. \ref{fig3}.
We will also prove that there always exists a parameter range in which the trade-off bound outperforms the divergence bound for any given pair of endpoints.

Fig. \ref{fig1} (c) displays a representative scenario where the endpoints are given and fixed, while also demonstrating universal behavior.
Eqs. \eqref{eq:divergencebound}, \eqref{eq:boundEPna} and \eqref{eq:newbound} jointly bound $\Sigma_\text{na}$, as represented by the shaded area.
The trade-off between dissipation and evolution time quantified by Eq. \eqref{eq:newbound} is clearly exemplified in this figure. 
It demonstrates that, when the endpoints are fixed, faster state transformations are  accompanied by larger amounts of dissipation.
As shown in the inset, each pseudo-coarse-grained two-state dynamics also has a speed limit, and the log-sum inequality implies that it is not greater than the speed limit of the $N$-state dynamics. 
The vertical asymptote of the two-state curve for $\Omega_3$ is exactly the speed limit for three states, and this is not a coincidence.
For any given pair of endpoints, there must be at least one two-state curve whose vertical asymptote coincides with the genuine speed limit for $N$ states.
Thus, there must  exist a parameter range in which the trade-off bound outperforms the divergence bound.

Fig. \ref{fig3} shows the entropy production $\Sigma$ and the corresponding divergence bound, activity bound and the new bound, as a function of the dimensionless time $\mathcal{W}t$ for a representative case, whose initial distribution and transition rate matrix (in arbitrary units) are given by
\begin{equation}
\begin{aligned}
&\boldsymbol{p}^\text{(i)} = \begin{bmatrix}
  1/3 &  1/3 &  1/3
\end{bmatrix}^\text{T} ,\\
&\boldsymbol{W} = \begin{bmatrix}
 -1.60 &  0.20 &  0.10\\
  0.70 & -0.80 &  0.70\\
  0.90 &  0.60 & -0.80
\end{bmatrix}
\end{aligned}.
\end{equation}
At each time instant $t$, we calculate the instantaneous distribution $\boldsymbol{p}(t) = \exp(\boldsymbol{W} t) \boldsymbol{p}^\text{(i)} $, and calculate the divergence bound and the new bound using corresponding equations by replacing $\boldsymbol{p}^\text{(f)}$ therein with $\boldsymbol{p}(t)$.
The time-averaged dynamical activity is calculated using Eq. (8) in Ref. \cite{Shiraishi2018}, with the upper bound of the integral set to $t$.
Hence, the time derivative of the divergence bound, the activity bound and the new bound cannot be regarded as entropy production rate, and are not guaranteed to be non-negative.
The divergence bound is initially loose but saturates as $t\to \infty$ as expected.
The activity bound has a much better performance than the divergence bound in the beginning but gradually loses its advantage.
The activity bound even decreases as the steady state is approached, showing the expected $1/\tau$ asymptotic behavior as previously mentioned.
The new bound with $\Omega_1=\{1\}$ has a performance that is almost all the time better than both the divergence bound and the activity bound.
As discussed above, in the long time limit the divergence bound's performance should be the best as it saturates.
For this special case, the trade-off bound is almost as good,
as can be numerically verified.
The steady state distribution is $\boldsymbol{p}^\text{(ss)}  = [0.086, 0.467, 0.447]^\text{T}$ , and the divergence bound is $0.2405$. The trade-off bound in the long-time limit is given by the divergence between $[1/3,2/3]^\text{T}$ and $[0.086,0.914]^\text{T}$, which is $0.2403$.

%
%
%

\emph{Proof of the speed limit.---}Let us prove that  Eq. \eqref{eq:speedlimit} gives a speed limit for relaxation processes in generic $N$-state systems, where the denominator is still the total rate $\mathcal{W}\equiv -\Tr \boldsymbol{W}$, i.e., the sum of all the positive rates.
First of all, let us divide \([0,\tau]\) into \(\mathcal{K} \gg 1\) intervals so a time sequence 
\(t_0=0< \ldots< t_k<\ldots< t_\mathcal{K}=\tau\) is obtained.
Consider an arbitrarily selected infinitesimal interval \([t_k,t_{k+1}]\), then the master equation  \eqref{eq:me} gives
\begin{equation}
  \delta p_m^{(k)} = (t_{k+1}-t_k) \sum_n W_{mn}p^{(k)}_n,  
\end{equation}
where \(\delta p_m^{(k)} =p_m^{(k+1)}-p_m^{(k)} \). Without loss of
generality, we assume the state $l$ satisfies
$\ln \big(p_{l}^{(k)} / p_{l}^{(k+1)}\big)=\max \limits_{n}\big\{\ln \big(p_{n}^{(k)} / p_{n}^{(k+1)}\big)\big\}$.
Therefore,
\begin{equation}
    t_{k+1}-t_k=\frac{\delta p_{l}^{(k)}}{W_{ll} p_{l}^{(k)}+\sum_{n \neq l} W_{l n} p_{n}^{(k)}}.
\end{equation}

If the sum of all positive rates is fixed, it is not hard to see that if
\(W_{ll}=-\mathcal{W}\) with all the other transition
rates vanishing, \(t_{k+1}-t_k\) reaches the minimum,
\begin{equation}
\label{eq:mintk}
\begin{aligned}
    t_{k+1}-t_k \geq \frac{\delta p_{l}^{(k)}}{(-\mathcal{W})p_{l}^{(k)}} &
    \approx  
    \frac{1}{\mathcal{W}} \ln \frac{p_l^{(k)}}{p_l^{(k+1)}}  \\
    & =\frac{1}{\mathcal{W}} \max _{1 \leq n \leq N}\left\{\ln \frac{p_{n}^{(k)}}{p_{n}^{(k+1)}}\right\},
\end{aligned}
\end{equation}
where $\approx$ is exact to first order in $\delta p_l^{(k)}/p_l^{(k)}$.
Technically speaking, the equality is not achievable in physical processes because \(W_{ll}=-\mathcal{W}\) is impractical.

Summing over $k$ results in 
\begin{equation}
\label{eq:wtau}
\mathcal{W} \tau  >   \sum_{k=0}^{\mathcal{K}-1} \max _{1\le n \le N}\left\{ \ln \frac{p_{n}^{(k)}}{p_{n}^{(k+1)}}  \right\}.
\end{equation}

The inequality that the maximum of sum is at most the sum of maxima gives
\begin{equation}
\begin{aligned}
 \sum_{k=0}^{\mathcal{K}-1} \max _{1\le n \le N}\left\{ \ln \frac{p_{n}^{(k)}}{p_{n}^{(k+1)}}  \right\}
  & > \max _{1\le n \le N} \left\{ \sum_{k=0}^{\mathcal{K}-1} \ln \frac{p_{n}^{(k)}}{p_{n}^{(k+1)}} \right\} \\
  & = \max\{ \ln ({p_n^\text{(i)}}/{p_n^\text{(f)})} \} .
\end{aligned}
\end{equation}
Combining with \eqref{eq:wtau} completes the proof.

\emph{Conclusion.---} 
We present a speed limit, a bound on the non-adiabatic entropy production and a trade-off relation between dissipation and evolution time for time-independent relaxation processes. 
Our result in Eq. \eqref{eq:speedlimit} resembles the quantum speed limit \eqref{eq:qsl} and indicates that the minimum transition time from an initial to a target state is constrained by the ratio of their $\infty$-R\'enyi divergence to the total rate. 
Eq. \eqref{eq:boundEPna} gives a lower bound on the non-adiabatic entropy production  in terms of $1$-R\'enyi divergence.
Furthermore,   Eq. \eqref{eq:newbound}, implicitly in the form of \eqref{eq:general}, reveals a new trade-off between dissipation and time that surpasses the divergence bound for certain parameters. Given successful quantum extensions of both divergence and activity bounds \cite{vanvu2021, vanvu2021a, vanvu2022,vanvu2023}, we anticipate that our results can also be generalized to quantum settings.

\emph{Acknowledgment.---}
We thank Naoto Shiraishi and anonymous referees for valuable comments and discussions.


\begin{thebibliography}{70}%
\makeatletter
\providecommand \@ifxundefined [1]{%
 \@ifx{#1\undefined}
}%
\providecommand \@ifnum [1]{%
 \ifnum #1\expandafter \@firstoftwo
 \else \expandafter \@secondoftwo
 \fi
}%
\providecommand \@ifx [1]{%
 \ifx #1\expandafter \@firstoftwo
 \else \expandafter \@secondoftwo
 \fi
}%
\providecommand \natexlab [1]{#1}%
\providecommand \enquote  [1]{``#1''}%
\providecommand \bibnamefont  [1]{#1}%
\providecommand \bibfnamefont [1]{#1}%
\providecommand \citenamefont [1]{#1}%
\providecommand \href@noop [0]{\@secondoftwo}%
\providecommand \href [0]{\begingroup \@sanitize@url \@href}%
\providecommand \@href[1]{\@@startlink{#1}\@@href}%
\providecommand \@@href[1]{\endgroup#1\@@endlink}%
\providecommand \@sanitize@url [0]{\catcode `\\12\catcode `\$12\catcode
  `\&12\catcode `\#12\catcode `\^12\catcode `\_12\catcode `\%12\relax}%
\providecommand \@@startlink[1]{}%
\providecommand \@@endlink[0]{}%
\providecommand \url  [0]{\begingroup\@sanitize@url \@url }%
\providecommand \@url [1]{\endgroup\@href {#1}{\urlprefix }}%
\providecommand \urlprefix  [0]{URL }%
\providecommand \Eprint [0]{\href }%
\providecommand \doibase [0]{https://doi.org/}%
\providecommand \selectlanguage [0]{\@gobble}%
\providecommand \bibinfo  [0]{\@secondoftwo}%
\providecommand \bibfield  [0]{\@secondoftwo}%
\providecommand \translation [1]{[#1]}%
\providecommand \BibitemOpen [0]{}%
\providecommand \bibitemStop [0]{}%
\providecommand \bibitemNoStop [0]{.\EOS\space}%
\providecommand \EOS [0]{\spacefactor3000\relax}%
\providecommand \BibitemShut  [1]{\csname bibitem#1\endcsname}%
\let\auto@bib@innerbib\@empty
\bibitem [{\citenamefont {Glaser}\ \emph {et~al.}(2015)\citenamefont {Glaser},
  \citenamefont {Boscain}, \citenamefont {Calarco}, \citenamefont {Koch},
  \citenamefont {K{\"{o}}ckenberger}, \citenamefont {Kosloff}, \citenamefont
  {Kuprov}, \citenamefont {Luy}, \citenamefont {Schirmer}, \citenamefont
  {Schulte-Herbr{\"{u}}ggen}, \citenamefont {Sugny},\ and\ \citenamefont
  {Wilhelm}}]{Glaser2015}%
  \BibitemOpen
  \bibfield  {author} {\bibinfo {author} {\bibfnamefont {S.~J.}\ \bibnamefont
  {Glaser}}, \bibinfo {author} {\bibfnamefont {U.}~\bibnamefont {Boscain}},
  \bibinfo {author} {\bibfnamefont {T.}~\bibnamefont {Calarco}}, \bibinfo
  {author} {\bibfnamefont {C.~P.}\ \bibnamefont {Koch}}, \bibinfo {author}
  {\bibfnamefont {W.}~\bibnamefont {K{\"{o}}ckenberger}}, \bibinfo {author}
  {\bibfnamefont {R.}~\bibnamefont {Kosloff}}, \bibinfo {author} {\bibfnamefont
  {I.}~\bibnamefont {Kuprov}}, \bibinfo {author} {\bibfnamefont
  {B.}~\bibnamefont {Luy}}, \bibinfo {author} {\bibfnamefont {S.}~\bibnamefont
  {Schirmer}}, \bibinfo {author} {\bibfnamefont {T.}~\bibnamefont
  {Schulte-Herbr{\"{u}}ggen}}, \bibinfo {author} {\bibfnamefont
  {D.}~\bibnamefont {Sugny}},\ and\ \bibinfo {author} {\bibfnamefont {F.~K.}\
  \bibnamefont {Wilhelm}},\ }\bibfield  {title} {\bibinfo {title} {{Training
  Schr{\"{o}}dinger's cat: quantum optimal control}},\ }\href
  {https://doi.org/10.1140/epjd/e2015-60464-1} {\bibfield  {journal} {\bibinfo
  {journal} {Eur. Phys. J. D}\ }\textbf {\bibinfo {volume} {69}},\ \bibinfo
  {pages} {279} (\bibinfo {year} {2015})}\BibitemShut {NoStop}%
\bibitem [{\citenamefont {Deffner}\ and\ \citenamefont
  {Bonan{\c{c}}a}(2020)}]{Deffner2020}%
  \BibitemOpen
  \bibfield  {author} {\bibinfo {author} {\bibfnamefont {S.}~\bibnamefont
  {Deffner}}\ and\ \bibinfo {author} {\bibfnamefont {M.~V.~S.}\ \bibnamefont
  {Bonan{\c{c}}a}},\ }\bibfield  {title} {\bibinfo {title} {{Thermodynamic
  control —An old paradigm with new applications}},\ }\href
  {https://doi.org/10.1209/0295-5075/131/20001} {\bibfield  {journal} {\bibinfo
   {journal} {Europhys. Lett.}\ }\textbf {\bibinfo {volume} {131}},\ \bibinfo
  {pages} {20001} (\bibinfo {year} {2020})}\BibitemShut {NoStop}%
\bibitem [{\citenamefont {Koch}\ \emph {et~al.}(2022)\citenamefont {Koch},
  \citenamefont {Boscain}, \citenamefont {Calarco}, \citenamefont {Dirr},
  \citenamefont {Filipp}, \citenamefont {Glaser}, \citenamefont {Kosloff},
  \citenamefont {Montangero}, \citenamefont {Schulte-Herbr{\"{u}}ggen},
  \citenamefont {Sugny},\ and\ \citenamefont {Wilhelm}}]{Koch2022}%
  \BibitemOpen
  \bibfield  {author} {\bibinfo {author} {\bibfnamefont {C.~P.}\ \bibnamefont
  {Koch}}, \bibinfo {author} {\bibfnamefont {U.}~\bibnamefont {Boscain}},
  \bibinfo {author} {\bibfnamefont {T.}~\bibnamefont {Calarco}}, \bibinfo
  {author} {\bibfnamefont {G.}~\bibnamefont {Dirr}}, \bibinfo {author}
  {\bibfnamefont {S.}~\bibnamefont {Filipp}}, \bibinfo {author} {\bibfnamefont
  {S.~J.}\ \bibnamefont {Glaser}}, \bibinfo {author} {\bibfnamefont
  {R.}~\bibnamefont {Kosloff}}, \bibinfo {author} {\bibfnamefont
  {S.}~\bibnamefont {Montangero}}, \bibinfo {author} {\bibfnamefont
  {T.}~\bibnamefont {Schulte-Herbr{\"{u}}ggen}}, \bibinfo {author}
  {\bibfnamefont {D.}~\bibnamefont {Sugny}},\ and\ \bibinfo {author}
  {\bibfnamefont {F.~K.}\ \bibnamefont {Wilhelm}},\ }\bibfield  {title}
  {\bibinfo {title} {{Quantum optimal control in quantum technologies.
  Strategic report on current status, visions and goals for research in
  Europe}},\ }\href {https://doi.org/10.1140/epjqt/s40507-022-00138-x}
  {\bibfield  {journal} {\bibinfo  {journal} {EPJ Quantum Technol.}\ }\textbf
  {\bibinfo {volume} {9}},\ \bibinfo {pages} {19} (\bibinfo {year}
  {2022})}\BibitemShut {NoStop}%
\bibitem [{\citenamefont {Blaber}\ and\ \citenamefont
  {Sivak}(2023)}]{blaber2023}%
  \BibitemOpen
  \bibfield  {author} {\bibinfo {author} {\bibfnamefont {S.}~\bibnamefont
  {Blaber}}\ and\ \bibinfo {author} {\bibfnamefont {D.~A.}\ \bibnamefont
  {Sivak}},\ }\bibfield  {title} {\bibinfo {title} {Optimal control in
  stochastic thermodynamics},\ }\href
  {https://doi.org/10.1088/2399-6528/acbf04} {\bibfield  {journal} {\bibinfo
  {journal} {J. Phys. Commun.}\ }\textbf {\bibinfo {volume} {7}},\ \bibinfo
  {pages} {033001} (\bibinfo {year} {2023})}\BibitemShut {NoStop}%
\bibitem [{\citenamefont {Villani}\ \emph {et~al.}(2009)\citenamefont {Villani}
  \emph {et~al.}}]{villani2009optimal}%
  \BibitemOpen
  \bibfield  {author} {\bibinfo {author} {\bibfnamefont {C.}~\bibnamefont
  {Villani}} \emph {et~al.},\ }\href@noop {} {\emph {\bibinfo {title} {Optimal
  transport: old and new}}},\ Vol.\ \bibinfo {volume} {338}\ (\bibinfo
  {publisher} {Springer},\ \bibinfo {year} {2009})\BibitemShut {NoStop}%
\bibitem [{\citenamefont {Van~Vu}\ and\ \citenamefont
  {Saito}(2023)}]{vanvu2023}%
  \BibitemOpen
  \bibfield  {author} {\bibinfo {author} {\bibfnamefont {T.}~\bibnamefont
  {Van~Vu}}\ and\ \bibinfo {author} {\bibfnamefont {K.}~\bibnamefont {Saito}},\
  }\bibfield  {title} {\bibinfo {title} {Thermodynamic {{Unification}} of
  {{Optimal Transport}}: {{Thermodynamic Uncertainty Relation}}, {{Minimum
  Dissipation}}, and {{Thermodynamic Speed Limits}}},\ }\href
  {https://doi.org/10.1103/PhysRevX.13.011013} {\bibfield  {journal} {\bibinfo
  {journal} {Phys. Rev. X}\ }\textbf {\bibinfo {volume} {13}},\ \bibinfo
  {pages} {011013} (\bibinfo {year} {2023})}\BibitemShut {NoStop}%
\bibitem [{\citenamefont {Gong}\ and\ \citenamefont
  {Hamazaki}(2022)}]{gong2022}%
  \BibitemOpen
  \bibfield  {author} {\bibinfo {author} {\bibfnamefont {Z.}~\bibnamefont
  {Gong}}\ and\ \bibinfo {author} {\bibfnamefont {R.}~\bibnamefont
  {Hamazaki}},\ }\bibfield  {title} {\bibinfo {title} {Bounds in
  {{Nonequilibrium Quantum Dynamics}}},\ }\href
  {https://doi.org/10.1142/S0217979222300079} {\bibfield  {journal} {\bibinfo
  {journal} {Int. J. Mod. Phys. B}\ }\textbf {\bibinfo {volume} {36}},\
  \bibinfo {pages} {2230007} (\bibinfo {year} {2022})}\BibitemShut {NoStop}%
\bibitem [{\citenamefont {Mandelstam}\ and\ \citenamefont
  {Tamm}(1945)}]{Mandelstam1945}%
  \BibitemOpen
  \bibfield  {author} {\bibinfo {author} {\bibfnamefont {L.}~\bibnamefont
  {Mandelstam}}\ and\ \bibinfo {author} {\bibfnamefont {I.}~\bibnamefont
  {Tamm}},\ }\bibfield  {title} {\bibinfo {title} {The uncertainty relation
  between energy and time in nonrelativistic quantum mechanics},\ }\href
  {https://link.springer.com/chapter/10.1007/978-3-642-74626-0_8} {\bibfield
  {journal} {\bibinfo  {journal} {J. Phys. (Moscow)}\ }\textbf {\bibinfo
  {volume} {9}},\ \bibinfo {pages} {249} (\bibinfo {year} {1945})}\BibitemShut
  {NoStop}%
\bibitem [{\citenamefont {Margolus}\ and\ \citenamefont
  {Levitin}(1998)}]{Margolus1998}%
  \BibitemOpen
  \bibfield  {author} {\bibinfo {author} {\bibfnamefont {N.}~\bibnamefont
  {Margolus}}\ and\ \bibinfo {author} {\bibfnamefont {L.~B.}\ \bibnamefont
  {Levitin}},\ }\bibfield  {title} {\bibinfo {title} {{The maximum speed of
  dynamical evolution}},\ }\href
  {https://doi.org/10.1016/S0167-2789(98)00054-2} {\bibfield  {journal}
  {\bibinfo  {journal} {Physica D}\ }\textbf {\bibinfo {volume} {120}},\
  \bibinfo {pages} {188} (\bibinfo {year} {1998})}\BibitemShut {NoStop}%
\bibitem [{\citenamefont {Deffner}\ and\ \citenamefont
  {Campbell}(2017)}]{Deffner2017}%
  \BibitemOpen
  \bibfield  {author} {\bibinfo {author} {\bibfnamefont {S.}~\bibnamefont
  {Deffner}}\ and\ \bibinfo {author} {\bibfnamefont {S.}~\bibnamefont
  {Campbell}},\ }\bibfield  {title} {\bibinfo {title} {{Quantum speed limits:
  from Heisenberg's uncertainty principle to optimal quantum control}},\ }\href
  {https://doi.org/10.1088/1751-8121/aa86c6} {\bibfield  {journal} {\bibinfo
  {journal} {J. Phys. A: Math. Theor.}\ }\textbf {\bibinfo {volume} {50}},\
  \bibinfo {pages} {453001} (\bibinfo {year} {2017})}\BibitemShut {NoStop}%
\bibitem [{\citenamefont {Giovannetti}\ \emph {et~al.}(2003)\citenamefont
  {Giovannetti}, \citenamefont {Lloyd},\ and\ \citenamefont
  {Maccone}}]{giovannetti2003}%
  \BibitemOpen
  \bibfield  {author} {\bibinfo {author} {\bibfnamefont {V.}~\bibnamefont
  {Giovannetti}}, \bibinfo {author} {\bibfnamefont {S.}~\bibnamefont {Lloyd}},\
  and\ \bibinfo {author} {\bibfnamefont {L.}~\bibnamefont {Maccone}},\
  }\bibfield  {title} {\bibinfo {title} {Quantum limits to dynamical
  evolution},\ }\href {https://doi.org/10.1103/PhysRevA.67.052109} {\bibfield
  {journal} {\bibinfo  {journal} {Phys. Rev. A}\ }\textbf {\bibinfo {volume}
  {67}},\ \bibinfo {pages} {052109} (\bibinfo {year} {2003})}\BibitemShut
  {NoStop}%
\bibitem [{\citenamefont {Giovannetti}\ \emph {et~al.}(2004)\citenamefont
  {Giovannetti}, \citenamefont {Lloyd},\ and\ \citenamefont
  {Maccone}}]{giovannetti2004}%
  \BibitemOpen
  \bibfield  {author} {\bibinfo {author} {\bibfnamefont {V.}~\bibnamefont
  {Giovannetti}}, \bibinfo {author} {\bibfnamefont {S.}~\bibnamefont {Lloyd}},\
  and\ \bibinfo {author} {\bibfnamefont {L.}~\bibnamefont {Maccone}},\
  }\bibfield  {title} {\bibinfo {title} {The speed limit of quantum unitary
  evolution},\ }\href {https://doi.org/10.1088/1464-4266/6/8/028} {\bibfield
  {journal} {\bibinfo  {journal} {J. Opt. B: Quantum Semiclass. Opt.}\ }\textbf
  {\bibinfo {volume} {6}},\ \bibinfo {pages} {S807} (\bibinfo {year}
  {2004})}\BibitemShut {NoStop}%
\bibitem [{\citenamefont {Shanahan}\ \emph {et~al.}(2018)\citenamefont
  {Shanahan}, \citenamefont {Chenu}, \citenamefont {Margolus},\ and\
  \citenamefont {{del Campo}}}]{shanahan2018}%
  \BibitemOpen
  \bibfield  {author} {\bibinfo {author} {\bibfnamefont {B.}~\bibnamefont
  {Shanahan}}, \bibinfo {author} {\bibfnamefont {A.}~\bibnamefont {Chenu}},
  \bibinfo {author} {\bibfnamefont {N.}~\bibnamefont {Margolus}},\ and\
  \bibinfo {author} {\bibfnamefont {A.}~\bibnamefont {{del Campo}}},\
  }\bibfield  {title} {\bibinfo {title} {Quantum {{Speed Limits}} across the
  {{Quantum-to-Classical Transition}}},\ }\href
  {https://doi.org/10.1103/PhysRevLett.120.070401} {\bibfield  {journal}
  {\bibinfo  {journal} {Phys. Rev. Lett.}\ }\textbf {\bibinfo {volume} {120}},\
  \bibinfo {pages} {070401} (\bibinfo {year} {2018})}\BibitemShut {NoStop}%
\bibitem [{\citenamefont {Okuyama}\ and\ \citenamefont
  {Ohzeki}(2018)}]{Okuyama2018}%
  \BibitemOpen
  \bibfield  {author} {\bibinfo {author} {\bibfnamefont {M.}~\bibnamefont
  {Okuyama}}\ and\ \bibinfo {author} {\bibfnamefont {M.}~\bibnamefont
  {Ohzeki}},\ }\bibfield  {title} {\bibinfo {title} {{Quantum Speed Limit is
  Not Quantum}},\ }\href {https://doi.org/10.1103/PhysRevLett.120.070402}
  {\bibfield  {journal} {\bibinfo  {journal} {Phys. Rev. Lett.}\ }\textbf
  {\bibinfo {volume} {120}},\ \bibinfo {pages} {070402} (\bibinfo {year}
  {2018})}\BibitemShut {NoStop}%
\bibitem [{\citenamefont {Shiraishi}\ \emph {et~al.}(2018)\citenamefont
  {Shiraishi}, \citenamefont {Funo},\ and\ \citenamefont
  {Saito}}]{Shiraishi2018}%
  \BibitemOpen
  \bibfield  {author} {\bibinfo {author} {\bibfnamefont {N.}~\bibnamefont
  {Shiraishi}}, \bibinfo {author} {\bibfnamefont {K.}~\bibnamefont {Funo}},\
  and\ \bibinfo {author} {\bibfnamefont {K.}~\bibnamefont {Saito}},\ }\bibfield
   {title} {\bibinfo {title} {{Speed Limit for Classical Stochastic
  Processes}},\ }\href {https://doi.org/10.1103/PhysRevLett.121.070601}
  {\bibfield  {journal} {\bibinfo  {journal} {Phys. Rev. Lett.}\ }\textbf
  {\bibinfo {volume} {121}},\ \bibinfo {pages} {070601} (\bibinfo {year}
  {2018})}\BibitemShut {NoStop}%
\bibitem [{\citenamefont {Vo}\ \emph {et~al.}(2020)\citenamefont {Vo},
  \citenamefont {{Van Vu}},\ and\ \citenamefont {Hasegawa}}]{Vo2020}%
  \BibitemOpen
  \bibfield  {author} {\bibinfo {author} {\bibfnamefont {V.~T.}\ \bibnamefont
  {Vo}}, \bibinfo {author} {\bibfnamefont {T.}~\bibnamefont {{Van Vu}}},\ and\
  \bibinfo {author} {\bibfnamefont {Y.}~\bibnamefont {Hasegawa}},\ }\bibfield
  {title} {\bibinfo {title} {{Unified approach to classical speed limit and
  thermodynamic uncertainty relation}},\ }\href
  {https://doi.org/10.1103/PhysRevE.102.062132} {\bibfield  {journal} {\bibinfo
   {journal} {Phys. Rev. E}\ }\textbf {\bibinfo {volume} {102}},\ \bibinfo
  {pages} {062132} (\bibinfo {year} {2020})}\BibitemShut {NoStop}%
\bibitem [{\citenamefont {Lee}\ \emph {et~al.}(2022)\citenamefont {Lee},
  \citenamefont {Lee}, \citenamefont {Kwon},\ and\ \citenamefont
  {Park}}]{Lee2022}%
  \BibitemOpen
  \bibfield  {author} {\bibinfo {author} {\bibfnamefont {J.~S.}\ \bibnamefont
  {Lee}}, \bibinfo {author} {\bibfnamefont {S.}~\bibnamefont {Lee}}, \bibinfo
  {author} {\bibfnamefont {H.}~\bibnamefont {Kwon}},\ and\ \bibinfo {author}
  {\bibfnamefont {H.}~\bibnamefont {Park}},\ }\bibfield  {title} {\bibinfo
  {title} {{Speed Limit for a Highly Irreversible Process and Tight Finite-Time
  Landauer's Bound}},\ }\href {https://doi.org/10.1103/PhysRevLett.129.120603}
  {\bibfield  {journal} {\bibinfo  {journal} {Phys. Rev. Lett.}\ }\textbf
  {\bibinfo {volume} {129}},\ \bibinfo {pages} {120603} (\bibinfo {year}
  {2022})}\BibitemShut {NoStop}%
\bibitem [{Note1()}]{Note1}%
  \BibitemOpen
  \bibinfo {note} {There are other speed limits that are not expressed in terms
  of entropy production. For example, the relevant quantity could be the rate
  of change in the information content \cite {nicholson2020,garcia-pintos2022},
  the entropy flux \cite {falasco2020}, or the dynamical activity \cite
  {dechant2022a}.}\BibitemShut {Stop}%
\bibitem [{\citenamefont {Hatano}\ and\ \citenamefont
  {Sasa}(2001)}]{Hatano2001}%
  \BibitemOpen
  \bibfield  {author} {\bibinfo {author} {\bibfnamefont {T.}~\bibnamefont
  {Hatano}}\ and\ \bibinfo {author} {\bibfnamefont {S.-i.}\ \bibnamefont
  {Sasa}},\ }\bibfield  {title} {\bibinfo {title} {{Steady-State Thermodynamics
  of Langevin Systems}},\ }\href {https://doi.org/10.1103/PhysRevLett.86.3463}
  {\bibfield  {journal} {\bibinfo  {journal} {Phys. Rev. Lett.}\ }\textbf
  {\bibinfo {volume} {86}},\ \bibinfo {pages} {3463} (\bibinfo {year}
  {2001})}\BibitemShut {NoStop}%
\bibitem [{\citenamefont {Esposito}\ and\ \citenamefont {{Van den
  Broeck}}(2010)}]{esposito2010}%
  \BibitemOpen
  \bibfield  {author} {\bibinfo {author} {\bibfnamefont {M.}~\bibnamefont
  {Esposito}}\ and\ \bibinfo {author} {\bibfnamefont {C.}~\bibnamefont {{Van
  den Broeck}}},\ }\bibfield  {title} {\bibinfo {title} {Three {{Detailed
  Fluctuation Theorems}}},\ }\href
  {https://doi.org/10.1103/PhysRevLett.104.090601} {\bibfield  {journal}
  {\bibinfo  {journal} {Phys. Rev. Lett.}\ }\textbf {\bibinfo {volume} {104}},\
  \bibinfo {pages} {090601} (\bibinfo {year} {2010})}\BibitemShut {NoStop}%
\bibitem [{\citenamefont {Maes}(2020)}]{Maes2020}%
  \BibitemOpen
  \bibfield  {author} {\bibinfo {author} {\bibfnamefont {C.}~\bibnamefont
  {Maes}},\ }\bibfield  {title} {\bibinfo {title} {{Frenesy: Time-symmetric
  dynamical activity in nonequilibria}},\ }\href
  {https://doi.org/10.1016/j.physrep.2020.01.002} {\bibfield  {journal}
  {\bibinfo  {journal} {Phys. Rep.}\ }\textbf {\bibinfo {volume} {850}},\
  \bibinfo {pages} {1} (\bibinfo {year} {2020})}\BibitemShut {NoStop}%
\bibitem [{\citenamefont {Sinitsyn}\ and\ \citenamefont
  {Nemenman}(2007)}]{Sinitsyn2007}%
  \BibitemOpen
  \bibfield  {author} {\bibinfo {author} {\bibfnamefont {N.~A.}\ \bibnamefont
  {Sinitsyn}}\ and\ \bibinfo {author} {\bibfnamefont {I.}~\bibnamefont
  {Nemenman}},\ }\bibfield  {title} {\bibinfo {title} {{The Berry phase and the
  pump flux in stochastic chemical kinetics}},\ }\href
  {https://doi.org/10.1209/0295-5075/77/58001} {\bibfield  {journal} {\bibinfo
  {journal} {Europhys. Lett.}\ }\textbf {\bibinfo {volume} {77}},\ \bibinfo
  {pages} {58001} (\bibinfo {year} {2007})}\BibitemShut {NoStop}%
\bibitem [{\citenamefont {Sinitsyn}(2009)}]{Sinitsyn2009}%
  \BibitemOpen
  \bibfield  {author} {\bibinfo {author} {\bibfnamefont {N.~A.}\ \bibnamefont
  {Sinitsyn}},\ }\bibfield  {title} {\bibinfo {title} {{The stochastic pump
  effect and geometric phases in dissipative and stochastic systems}},\ }\href
  {https://doi.org/10.1088/1751-8113/42/19/193001} {\bibfield  {journal}
  {\bibinfo  {journal} {J. Phys. A: Math. Theor.}\ }\textbf {\bibinfo {volume}
  {42}},\ \bibinfo {pages} {193001} (\bibinfo {year} {2009})}\BibitemShut
  {NoStop}%
\bibitem [{\citenamefont {Ren}\ \emph {et~al.}(2010)\citenamefont {Ren},
  \citenamefont {H{\"{a}}nggi},\ and\ \citenamefont {Li}}]{Ren2010}%
  \BibitemOpen
  \bibfield  {author} {\bibinfo {author} {\bibfnamefont {J.}~\bibnamefont
  {Ren}}, \bibinfo {author} {\bibfnamefont {P.}~\bibnamefont {H{\"{a}}nggi}},\
  and\ \bibinfo {author} {\bibfnamefont {B.}~\bibnamefont {Li}},\ }\bibfield
  {title} {\bibinfo {title} {{Berry-Phase-Induced Heat Pumping and Its Impact
  on the Fluctuation Theorem}},\ }\href
  {https://doi.org/10.1103/PhysRevLett.104.170601} {\bibfield  {journal}
  {\bibinfo  {journal} {Phys. Rev. Lett.}\ }\textbf {\bibinfo {volume} {104}},\
  \bibinfo {pages} {170601} (\bibinfo {year} {2010})}\BibitemShut {NoStop}%
\bibitem [{\citenamefont {Sinitsyn}\ \emph {et~al.}(2011)\citenamefont
  {Sinitsyn}, \citenamefont {Akimov},\ and\ \citenamefont
  {Chernyak}}]{sinitsyn2011}%
  \BibitemOpen
  \bibfield  {author} {\bibinfo {author} {\bibfnamefont {N.~A.}\ \bibnamefont
  {Sinitsyn}}, \bibinfo {author} {\bibfnamefont {A.}~\bibnamefont {Akimov}},\
  and\ \bibinfo {author} {\bibfnamefont {V.~Y.}\ \bibnamefont {Chernyak}},\
  }\bibfield  {title} {\bibinfo {title} {Supersymmetry and fluctuation
  relations for currents in closed networks},\ }\href
  {https://doi.org/10.1103/PhysRevE.83.021107} {\bibfield  {journal} {\bibinfo
  {journal} {Phys. Rev. E}\ }\textbf {\bibinfo {volume} {83}},\ \bibinfo
  {pages} {021107} (\bibinfo {year} {2011})}\BibitemShut {NoStop}%
\bibitem [{\citenamefont {Gu}\ \emph {et~al.}(2018)\citenamefont {Gu},
  \citenamefont {Li}, \citenamefont {Cheng},\ and\ \citenamefont
  {Zhang}}]{Gu2017}%
  \BibitemOpen
  \bibfield  {author} {\bibinfo {author} {\bibfnamefont {J.}~\bibnamefont
  {Gu}}, \bibinfo {author} {\bibfnamefont {X.-G.}\ \bibnamefont {Li}}, \bibinfo
  {author} {\bibfnamefont {H.-P.}\ \bibnamefont {Cheng}},\ and\ \bibinfo
  {author} {\bibfnamefont {X.-G.}\ \bibnamefont {Zhang}},\ }\bibfield  {title}
  {\bibinfo {title} {{Adiabatic Spin Pump through a Molecular Antiferromagnet
  Ce 3 Mn 8 III}},\ }\href {https://doi.org/10.1021/acs.jpcc.7b11387}
  {\bibfield  {journal} {\bibinfo  {journal} {J. Phys. Chem. C}\ }\textbf
  {\bibinfo {volume} {122}},\ \bibinfo {pages} {1422} (\bibinfo {year}
  {2018})}\BibitemShut {NoStop}%
\bibitem [{\citenamefont {Kirkwood}(1946)}]{Kirkwood1946}%
  \BibitemOpen
  \bibfield  {author} {\bibinfo {author} {\bibfnamefont {J.~G.}\ \bibnamefont
  {Kirkwood}},\ }\bibfield  {title} {\bibinfo {title} {{The Statistical
  Mechanical Theory of Transport Processes I. General Theory}},\ }\href
  {https://doi.org/10.1063/1.1724117} {\bibfield  {journal} {\bibinfo
  {journal} {J. Chem. Phys.}\ }\textbf {\bibinfo {volume} {14}},\ \bibinfo
  {pages} {180} (\bibinfo {year} {1946})}\BibitemShut {NoStop}%
\bibitem [{\citenamefont {Weinhold}(1975)}]{Weinhold1975}%
  \BibitemOpen
  \bibfield  {author} {\bibinfo {author} {\bibfnamefont {F.}~\bibnamefont
  {Weinhold}},\ }\bibfield  {title} {\bibinfo {title} {{Metric geometry of
  equilibrium thermodynamics}},\ }\href {https://doi.org/10.1063/1.431689}
  {\bibfield  {journal} {\bibinfo  {journal} {J. Chem. Phys.}\ }\textbf
  {\bibinfo {volume} {63}},\ \bibinfo {pages} {2479} (\bibinfo {year}
  {1975})}\BibitemShut {NoStop}%
\bibitem [{\citenamefont {Ruppeiner}(1979)}]{Ruppeiner1979}%
  \BibitemOpen
  \bibfield  {author} {\bibinfo {author} {\bibfnamefont {G.}~\bibnamefont
  {Ruppeiner}},\ }\bibfield  {title} {\bibinfo {title} {{Thermodynamics: A
  Riemannian geometric model}},\ }\href
  {https://doi.org/10.1103/PhysRevA.20.1608} {\bibfield  {journal} {\bibinfo
  {journal} {Phys. Rev. A}\ }\textbf {\bibinfo {volume} {20}},\ \bibinfo
  {pages} {1608} (\bibinfo {year} {1979})}\BibitemShut {NoStop}%
\bibitem [{\citenamefont {Salamon}\ and\ \citenamefont
  {Berry}(1983)}]{Salamon1983}%
  \BibitemOpen
  \bibfield  {author} {\bibinfo {author} {\bibfnamefont {P.}~\bibnamefont
  {Salamon}}\ and\ \bibinfo {author} {\bibfnamefont {R.~S.}\ \bibnamefont
  {Berry}},\ }\bibfield  {title} {\bibinfo {title} {{Thermodynamic length and
  dissipated availability}},\ }\href
  {https://doi.org/10.1103/PhysRevLett.51.1127} {\bibfield  {journal} {\bibinfo
   {journal} {Phys. Rev. Lett.}\ }\textbf {\bibinfo {volume} {51}},\ \bibinfo
  {pages} {1127} (\bibinfo {year} {1983})}\BibitemShut {NoStop}%
\bibitem [{\citenamefont {Janyszek}\ and\ \citenamefont
  {Mrugal/a}(1989)}]{Janyszek1989}%
  \BibitemOpen
  \bibfield  {author} {\bibinfo {author} {\bibfnamefont {H.}~\bibnamefont
  {Janyszek}}\ and\ \bibinfo {author} {\bibfnamefont {R.}~\bibnamefont
  {Mrugal/a}},\ }\bibfield  {title} {\bibinfo {title} {{Riemannian geometry and
  the thermodynamics of model magnetic systems}},\ }\href
  {https://doi.org/10.1103/PhysRevA.39.6515} {\bibfield  {journal} {\bibinfo
  {journal} {Phys. Rev. A}\ }\textbf {\bibinfo {volume} {39}},\ \bibinfo
  {pages} {6515} (\bibinfo {year} {1989})}\BibitemShut {NoStop}%
\bibitem [{\citenamefont {Brody}\ and\ \citenamefont
  {Rivier}(1995)}]{Brody1995}%
  \BibitemOpen
  \bibfield  {author} {\bibinfo {author} {\bibfnamefont {D.}~\bibnamefont
  {Brody}}\ and\ \bibinfo {author} {\bibfnamefont {N.}~\bibnamefont {Rivier}},\
  }\bibfield  {title} {\bibinfo {title} {{Geometrical aspects of statistical
  mechanics}},\ }\href {https://doi.org/10.1103/PhysRevE.51.1006} {\bibfield
  {journal} {\bibinfo  {journal} {Phys. Rev. E}\ }\textbf {\bibinfo {volume}
  {51}},\ \bibinfo {pages} {1006} (\bibinfo {year} {1995})}\BibitemShut
  {NoStop}%
\bibitem [{\citenamefont {Ruppeiner}(1995)}]{Ruppeiner1995}%
  \BibitemOpen
  \bibfield  {author} {\bibinfo {author} {\bibfnamefont {G.}~\bibnamefont
  {Ruppeiner}},\ }\bibfield  {title} {\bibinfo {title} {{Riemannian geometry in
  thermodynamic fluctuation theory}},\ }\href
  {https://doi.org/10.1103/RevModPhys.67.605} {\bibfield  {journal} {\bibinfo
  {journal} {Rev. Mod. Phys.}\ }\textbf {\bibinfo {volume} {67}},\ \bibinfo
  {pages} {605} (\bibinfo {year} {1995})}\BibitemShut {NoStop}%
\bibitem [{\citenamefont {Sivak}\ and\ \citenamefont
  {Crooks}(2012)}]{Sivak2012}%
  \BibitemOpen
  \bibfield  {author} {\bibinfo {author} {\bibfnamefont {D.~A.}\ \bibnamefont
  {Sivak}}\ and\ \bibinfo {author} {\bibfnamefont {G.~E.}\ \bibnamefont
  {Crooks}},\ }\bibfield  {title} {\bibinfo {title} {{Thermodynamic Metrics and
  Optimal Paths}},\ }\href {https://doi.org/10.1103/PhysRevLett.108.190602}
  {\bibfield  {journal} {\bibinfo  {journal} {Phys. Rev. Lett.}\ }\textbf
  {\bibinfo {volume} {108}},\ \bibinfo {pages} {190602} (\bibinfo {year}
  {2012})}\BibitemShut {NoStop}%
\bibitem [{\citenamefont {Zulkowski}\ \emph {et~al.}(2012)\citenamefont
  {Zulkowski}, \citenamefont {Sivak}, \citenamefont {Crooks},\ and\
  \citenamefont {DeWeese}}]{Zulkowski2012}%
  \BibitemOpen
  \bibfield  {author} {\bibinfo {author} {\bibfnamefont {P.~R.}\ \bibnamefont
  {Zulkowski}}, \bibinfo {author} {\bibfnamefont {D.~A.}\ \bibnamefont
  {Sivak}}, \bibinfo {author} {\bibfnamefont {G.~E.}\ \bibnamefont {Crooks}},\
  and\ \bibinfo {author} {\bibfnamefont {M.~R.}\ \bibnamefont {DeWeese}},\
  }\bibfield  {title} {\bibinfo {title} {{Geometry of thermodynamic control}},\
  }\href {https://doi.org/10.1103/PhysRevE.86.041148} {\bibfield  {journal}
  {\bibinfo  {journal} {Phys. Rev. E}\ }\textbf {\bibinfo {volume} {86}},\
  \bibinfo {pages} {041148} (\bibinfo {year} {2012})}\BibitemShut {NoStop}%
\bibitem [{\citenamefont {Machta}(2015)}]{machta2015}%
  \BibitemOpen
  \bibfield  {author} {\bibinfo {author} {\bibfnamefont {B.~B.}\ \bibnamefont
  {Machta}},\ }\bibfield  {title} {\bibinfo {title} {Dissipation {{Bound}} for
  {{Thermodynamic Control}}},\ }\href
  {https://doi.org/10.1103/PhysRevLett.115.260603} {\bibfield  {journal}
  {\bibinfo  {journal} {Phys. Rev. Lett.}\ }\textbf {\bibinfo {volume} {115}},\
  \bibinfo {pages} {260603} (\bibinfo {year} {2015})}\BibitemShut {NoStop}%
\bibitem [{\citenamefont {Mandal}\ and\ \citenamefont
  {Jarzynski}(2016)}]{Mandal2016}%
  \BibitemOpen
  \bibfield  {author} {\bibinfo {author} {\bibfnamefont {D.}~\bibnamefont
  {Mandal}}\ and\ \bibinfo {author} {\bibfnamefont {C.}~\bibnamefont
  {Jarzynski}},\ }\bibfield  {title} {\bibinfo {title} {{Analysis of slow
  transitions between nonequilibrium steady states}},\ }\href
  {https://doi.org/10.1088/1742-5468/2016/06/063204} {\bibfield  {journal}
  {\bibinfo  {journal} {J. Stat. Mech: Theory Exp.}\ }\textbf {\bibinfo
  {volume} {2016}},\ \bibinfo {pages} {063204} (\bibinfo {year}
  {2016})}\BibitemShut {NoStop}%
\bibitem [{\citenamefont {Miller}\ \emph {et~al.}(2019)\citenamefont {Miller},
  \citenamefont {Scandi}, \citenamefont {Anders},\ and\ \citenamefont
  {Perarnau-Llobet}}]{Miller2019}%
  \BibitemOpen
  \bibfield  {author} {\bibinfo {author} {\bibfnamefont {H.~J.}\ \bibnamefont
  {Miller}}, \bibinfo {author} {\bibfnamefont {M.}~\bibnamefont {Scandi}},
  \bibinfo {author} {\bibfnamefont {J.}~\bibnamefont {Anders}},\ and\ \bibinfo
  {author} {\bibfnamefont {M.}~\bibnamefont {Perarnau-Llobet}},\ }\bibfield
  {title} {\bibinfo {title} {{Work Fluctuations in Slow Processes: Quantum
  Signatures and Optimal Control}},\ }\href
  {https://doi.org/10.1103/PhysRevLett.123.230603} {\bibfield  {journal}
  {\bibinfo  {journal} {Phys. Rev. Lett.}\ }\textbf {\bibinfo {volume} {123}},\
  \bibinfo {pages} {230603} (\bibinfo {year} {2019})}\BibitemShut {NoStop}%
\bibitem [{\citenamefont {Miller}\ and\ \citenamefont
  {Mehboudi}(2020)}]{Miller2020}%
  \BibitemOpen
  \bibfield  {author} {\bibinfo {author} {\bibfnamefont {H.~J.}\ \bibnamefont
  {Miller}}\ and\ \bibinfo {author} {\bibfnamefont {M.}~\bibnamefont
  {Mehboudi}},\ }\bibfield  {title} {\bibinfo {title} {{Geometry of Work
  Fluctuations versus Efficiency in Microscopic Thermal Machines}},\ }\href
  {https://doi.org/10.1103/PhysRevLett.125.260602} {\bibfield  {journal}
  {\bibinfo  {journal} {Phys. Rev. Lett.}\ }\textbf {\bibinfo {volume} {125}},\
  \bibinfo {pages} {260602} (\bibinfo {year} {2020})}\BibitemShut {NoStop}%
\bibitem [{\citenamefont {Scandi}\ and\ \citenamefont
  {Perarnau-Llobet}(2019)}]{Scandi2019}%
  \BibitemOpen
  \bibfield  {author} {\bibinfo {author} {\bibfnamefont {M.}~\bibnamefont
  {Scandi}}\ and\ \bibinfo {author} {\bibfnamefont {M.}~\bibnamefont
  {Perarnau-Llobet}},\ }\bibfield  {title} {\bibinfo {title} {{Thermodynamic
  length in open quantum systems}},\ }\href
  {https://doi.org/10.22331/q-2019-10-24-197} {\bibfield  {journal} {\bibinfo
  {journal} {Quantum}\ }\textbf {\bibinfo {volume} {3}},\ \bibinfo {pages}
  {197} (\bibinfo {year} {2019})}\BibitemShut {NoStop}%
\bibitem [{\citenamefont {Scandi}\ \emph {et~al.}(2020)\citenamefont {Scandi},
  \citenamefont {Miller}, \citenamefont {Anders},\ and\ \citenamefont
  {Perarnau-Llobet}}]{Scandi2020}%
  \BibitemOpen
  \bibfield  {author} {\bibinfo {author} {\bibfnamefont {M.}~\bibnamefont
  {Scandi}}, \bibinfo {author} {\bibfnamefont {H.~J.~D.}\ \bibnamefont
  {Miller}}, \bibinfo {author} {\bibfnamefont {J.}~\bibnamefont {Anders}},\
  and\ \bibinfo {author} {\bibfnamefont {M.}~\bibnamefont {Perarnau-Llobet}},\
  }\bibfield  {title} {\bibinfo {title} {{Quantum work statistics close to
  equilibrium}},\ }\href {https://doi.org/10.1103/PhysRevResearch.2.023377}
  {\bibfield  {journal} {\bibinfo  {journal} {Phys. Rev. Res.}\ }\textbf
  {\bibinfo {volume} {2}},\ \bibinfo {pages} {023377} (\bibinfo {year}
  {2020})}\BibitemShut {NoStop}%
\bibitem [{\citenamefont {Abiuso}\ \emph {et~al.}(2020)\citenamefont {Abiuso},
  \citenamefont {Miller}, \citenamefont {Perarnau-Llobet},\ and\ \citenamefont
  {Scandi}}]{Abiuso2020}%
  \BibitemOpen
  \bibfield  {author} {\bibinfo {author} {\bibfnamefont {P.}~\bibnamefont
  {Abiuso}}, \bibinfo {author} {\bibfnamefont {H.~J.~D.}\ \bibnamefont
  {Miller}}, \bibinfo {author} {\bibfnamefont {M.}~\bibnamefont
  {Perarnau-Llobet}},\ and\ \bibinfo {author} {\bibfnamefont {M.}~\bibnamefont
  {Scandi}},\ }\bibfield  {title} {\bibinfo {title} {{Geometric Optimisation of
  Quantum Thermodynamic Processes}},\ }\href
  {https://doi.org/10.3390/e22101076} {\bibfield  {journal} {\bibinfo
  {journal} {Entropy}\ }\textbf {\bibinfo {volume} {22}},\ \bibinfo {pages}
  {1076} (\bibinfo {year} {2020})}\BibitemShut {NoStop}%
\bibitem [{\citenamefont {Frim}\ and\ \citenamefont
  {DeWeese}(2022)}]{Frim2022}%
  \BibitemOpen
  \bibfield  {author} {\bibinfo {author} {\bibfnamefont {A.~G.}\ \bibnamefont
  {Frim}}\ and\ \bibinfo {author} {\bibfnamefont {M.~R.}\ \bibnamefont
  {DeWeese}},\ }\bibfield  {title} {\bibinfo {title} {{Geometric Bound on the
  Efficiency of Irreversible Thermodynamic Cycles}},\ }\href
  {https://doi.org/10.1103/PhysRevLett.128.230601} {\bibfield  {journal}
  {\bibinfo  {journal} {Phys. Rev. Lett.}\ }\textbf {\bibinfo {volume} {128}},\
  \bibinfo {pages} {230601} (\bibinfo {year} {2022})}\BibitemShut {NoStop}%
\bibitem [{\citenamefont {Gu}(2023)}]{gu2023}%
  \BibitemOpen
  \bibfield  {author} {\bibinfo {author} {\bibfnamefont {J.}~\bibnamefont
  {Gu}},\ }\bibfield  {title} {\bibinfo {title} {Work statistics in slow
  thermodynamic processes},\ }\href {https://doi.org/10.1063/5.0138405}
  {\bibfield  {journal} {\bibinfo  {journal} {J. Chem. Phys.}\ }\textbf
  {\bibinfo {volume} {158}},\ \bibinfo {pages} {074104} (\bibinfo {year}
  {2023})}\BibitemShut {NoStop}%
\bibitem [{\citenamefont {Shiraishi}\ and\ \citenamefont
  {Saito}(2019)}]{Shiraishi2019}%
  \BibitemOpen
  \bibfield  {author} {\bibinfo {author} {\bibfnamefont {N.}~\bibnamefont
  {Shiraishi}}\ and\ \bibinfo {author} {\bibfnamefont {K.}~\bibnamefont
  {Saito}},\ }\bibfield  {title} {\bibinfo {title} {{Information-Theoretical
  Bound of the Irreversibility in Thermal Relaxation Processes}},\ }\href
  {https://doi.org/10.1103/PhysRevLett.123.110603} {\bibfield  {journal}
  {\bibinfo  {journal} {Phys. Rev. Lett.}\ }\textbf {\bibinfo {volume} {123}},\
  \bibinfo {pages} {110603} (\bibinfo {year} {2019})}\BibitemShut {NoStop}%
\bibitem [{\citenamefont {van Erven}\ and\ \citenamefont
  {Harremoes}(2014)}]{VanErven2014}%
  \BibitemOpen
  \bibfield  {author} {\bibinfo {author} {\bibfnamefont {T.}~\bibnamefont {van
  Erven}}\ and\ \bibinfo {author} {\bibfnamefont {P.}~\bibnamefont
  {Harremoes}},\ }\bibfield  {title} {\bibinfo {title} {{R{\'{e}}nyi Divergence
  and Kullback-Leibler Divergence}},\ }\href
  {https://doi.org/10.1109/TIT.2014.2320500} {\bibfield  {journal} {\bibinfo
  {journal} {IEEE Trans. Inf. Theory}\ }\textbf {\bibinfo {volume} {60}},\
  \bibinfo {pages} {3797} (\bibinfo {year} {2014})}\BibitemShut {NoStop}%
\bibitem [{\citenamefont {Sagawa}(2022)}]{sagawa2022entropy}%
  \BibitemOpen
  \bibfield  {author} {\bibinfo {author} {\bibfnamefont {T.}~\bibnamefont
  {Sagawa}},\ }\href@noop {} {\emph {\bibinfo {title} {Entropy, Divergence, and
  Majorization in Classical and Quantum Thermodynamics}}},\ Vol.~\bibinfo
  {volume} {16}\ (\bibinfo  {publisher} {Springer Nature},\ \bibinfo {year}
  {2022})\BibitemShut {NoStop}%
\bibitem [{\citenamefont {Vedral}(2002)}]{Vedral2002}%
  \BibitemOpen
  \bibfield  {author} {\bibinfo {author} {\bibfnamefont {V.}~\bibnamefont
  {Vedral}},\ }\bibfield  {title} {\bibinfo {title} {{The role of relative
  entropy in quantum information theory}},\ }\href
  {https://doi.org/10.1103/RevModPhys.74.197} {\bibfield  {journal} {\bibinfo
  {journal} {Rev. Mod. Phys.}\ }\textbf {\bibinfo {volume} {74}},\ \bibinfo
  {pages} {197} (\bibinfo {year} {2002})}\BibitemShut {NoStop}%
\bibitem [{\citenamefont {Van~Kampen}(1992)}]{van1992}%
  \BibitemOpen
  \bibfield  {author} {\bibinfo {author} {\bibfnamefont {N.~G.}\ \bibnamefont
  {Van~Kampen}},\ }\href@noop {} {\emph {\bibinfo {title} {Stochastic processes
  in physics and chemistry}}},\ Vol.~\bibinfo {volume} {1}\ (\bibinfo
  {publisher} {Elsevier},\ \bibinfo {year} {1992})\BibitemShut {NoStop}%
\bibitem [{Note2()}]{Note2}%
  \BibitemOpen
  \bibinfo {note} {There exists a relevant speed limit for open quantum systems
  governed by a Markovian quantum master equation \cite {delcampo2013}.
  Surprisingly, the bound therein only involves the initial state and the
  dynamical map.}\BibitemShut {Stop}%
\bibitem [{\citenamefont {Bladt}\ and\ \citenamefont
  {S{\o}rensen}(2005)}]{bladt2005}%
  \BibitemOpen
  \bibfield  {author} {\bibinfo {author} {\bibfnamefont {M.}~\bibnamefont
  {Bladt}}\ and\ \bibinfo {author} {\bibfnamefont {M.}~\bibnamefont
  {S{\o}rensen}},\ }\bibfield  {title} {\bibinfo {title} {Statistical
  {{Inference}} for {{Discretely Observed Markov Jump Processes}}},\ }\href
  {https://doi.org/10.1111/j.1467-9868.2005.00508.x} {\bibfield  {journal}
  {\bibinfo  {journal} {J. R. Stat. Soc., B: Stat.}\ }\textbf {\bibinfo
  {volume} {67}},\ \bibinfo {pages} {395} (\bibinfo {year} {2005})}\BibitemShut
  {NoStop}%
\bibitem [{\citenamefont {{Van den Broeck C.}}(2013)}]{broeck2013}%
  \BibitemOpen
  \bibfield  {author} {\bibinfo {author} {\bibnamefont {{Van den Broeck C.}}},\
  }\bibfield  {title} {\bibinfo {title} {Stochastic thermodynamics: {{A}} brief
  introduction},\ }\href {https://doi.org/10.3254/978-1-61499-278-3-155}
  {\bibfield  {journal} {\bibinfo  {journal} {ENFI}\ }\textbf {\bibinfo
  {volume} {184}},\ \bibinfo {pages} {155} (\bibinfo {year}
  {2013})}\BibitemShut {NoStop}%
\bibitem [{Note3()}]{Note3}%
  \BibitemOpen
  \bibinfo {note} {See the Supplemental Material at [url] for derivation of Eq.
  \protect \textup {\hbox {\mathsurround \z@ \protect \normalfont
  (\ignorespaces \ref {eq:p1}\unskip \@@italiccorr )}} and the code to verify
  Eqs. \protect \textup {\hbox {\mathsurround \z@ \protect \normalfont
  (\ignorespaces \ref {eq:boundEPna}\unskip \@@italiccorr )}} and \protect
  \textup {\hbox {\mathsurround \z@ \protect \normalfont (\ignorespaces \ref
  {eq:newbound}\unskip \@@italiccorr )}}.}\BibitemShut {Stop}%
\bibitem [{\citenamefont {Horodecki}\ and\ \citenamefont
  {Oppenheim}(2013)}]{horodecki2013}%
  \BibitemOpen
  \bibfield  {author} {\bibinfo {author} {\bibfnamefont {M.}~\bibnamefont
  {Horodecki}}\ and\ \bibinfo {author} {\bibfnamefont {J.}~\bibnamefont
  {Oppenheim}},\ }\bibfield  {title} {\bibinfo {title} {Fundamental limitations
  for quantum and nanoscale thermodynamics},\ }\href
  {https://doi.org/10.1038/ncomms3059} {\bibfield  {journal} {\bibinfo
  {journal} {Nat. Commun.}\ }\textbf {\bibinfo {volume} {4}},\ \bibinfo {pages}
  {2059} (\bibinfo {year} {2013})}\BibitemShut {NoStop}%
\bibitem [{\citenamefont {{\AA}berg}(2013)}]{aberg2013}%
  \BibitemOpen
  \bibfield  {author} {\bibinfo {author} {\bibfnamefont {J.}~\bibnamefont
  {{\AA}berg}},\ }\bibfield  {title} {\bibinfo {title} {Truly work-like work
  extraction via a single-shot analysis},\ }\href
  {https://doi.org/10.1038/ncomms2712} {\bibfield  {journal} {\bibinfo
  {journal} {Nat. Commun.}\ }\textbf {\bibinfo {volume} {4}},\ \bibinfo {pages}
  {1925} (\bibinfo {year} {2013})}\BibitemShut {NoStop}%
\bibitem [{\citenamefont {Masanes}\ and\ \citenamefont
  {Oppenheim}(2017)}]{masanes2017}%
  \BibitemOpen
  \bibfield  {author} {\bibinfo {author} {\bibfnamefont {L.}~\bibnamefont
  {Masanes}}\ and\ \bibinfo {author} {\bibfnamefont {J.}~\bibnamefont
  {Oppenheim}},\ }\bibfield  {title} {\bibinfo {title} {A general derivation
  and quantification of the third law of thermodynamics},\ }\href
  {https://doi.org/10.1038/ncomms14538} {\bibfield  {journal} {\bibinfo
  {journal} {Nat. Commun.}\ }\textbf {\bibinfo {volume} {8}},\ \bibinfo {pages}
  {14538} (\bibinfo {year} {2017})}\BibitemShut {NoStop}%
\bibitem [{\citenamefont {Kolchinsky}\ \emph {et~al.}(2022)\citenamefont
  {Kolchinsky}, \citenamefont {Dechant}, \citenamefont {Yoshimura},\ and\
  \citenamefont {Ito}}]{kolchinsky2022}%
  \BibitemOpen
  \bibfield  {author} {\bibinfo {author} {\bibfnamefont {A.}~\bibnamefont
  {Kolchinsky}}, \bibinfo {author} {\bibfnamefont {A.}~\bibnamefont {Dechant}},
  \bibinfo {author} {\bibfnamefont {K.}~\bibnamefont {Yoshimura}},\ and\
  \bibinfo {author} {\bibfnamefont {S.}~\bibnamefont {Ito}},\ }\bibfield
  {title} {\bibinfo {title} {Information geometry of excess and housekeeping
  entropy production}} (\bibinfo {year} {2022})\BibitemShut {NoStop}%
\bibitem [{\citenamefont {Shiraishi}(2023)}]{shiraishi2023a}%
  \BibitemOpen
  \bibfield  {author} {\bibinfo {author} {\bibfnamefont {N.}~\bibnamefont
  {Shiraishi}},\ }\bibinfo {title} {An {{Introduction}} to {{Stochastic
  Thermodynamics}}: {{From Basic}} to {{Advanced}}}\ (\bibinfo  {publisher}
  {{Springer Nature Singapore}},\ \bibinfo {address} {{Singapore}},\ \bibinfo
  {year} {2023})\ Chap.~\bibinfo {chapter} {19}\BibitemShut {NoStop}%
\bibitem [{\citenamefont {Abramowitz}\ and\ \citenamefont
  {Stegun}(1965)}]{Abramowitz1965}%
  \BibitemOpen
  \bibinfo {editor} {\bibfnamefont {M.}~\bibnamefont {Abramowitz}}\ and\
  \bibinfo {editor} {\bibfnamefont {I.~A.}\ \bibnamefont {Stegun}},\ eds.,\
  \href@noop {} {\emph {\bibinfo {title} {Handbook of mathematical
  functions}}},\ Dover Books on Mathematics\ (\bibinfo  {publisher} {Dover
  Publications},\ \bibinfo {address} {Mineola, NY},\ \bibinfo {year}
  {1965})\BibitemShut {NoStop}%
\bibitem [{\citenamefont {Esposito}(2012)}]{esposito2012}%
  \BibitemOpen
  \bibfield  {author} {\bibinfo {author} {\bibfnamefont {M.}~\bibnamefont
  {Esposito}},\ }\bibfield  {title} {\bibinfo {title} {Stochastic
  thermodynamics under coarse-graining},\ }\href
  {https://doi.org/10.1103/PhysRevE.85.041125} {\bibfield  {journal} {\bibinfo
  {journal} {Phys. Rev. E}\ }\textbf {\bibinfo {volume} {85}},\ \bibinfo
  {pages} {041125} (\bibinfo {year} {2012})}\BibitemShut {NoStop}%
\bibitem [{\citenamefont {Zhen}\ \emph {et~al.}(2021)\citenamefont {Zhen},
  \citenamefont {Egloff}, \citenamefont {Modi},\ and\ \citenamefont
  {Dahlsten}}]{zhen2021}%
  \BibitemOpen
  \bibfield  {author} {\bibinfo {author} {\bibfnamefont {Y.-Z.}\ \bibnamefont
  {Zhen}}, \bibinfo {author} {\bibfnamefont {D.}~\bibnamefont {Egloff}},
  \bibinfo {author} {\bibfnamefont {K.}~\bibnamefont {Modi}},\ and\ \bibinfo
  {author} {\bibfnamefont {O.}~\bibnamefont {Dahlsten}},\ }\bibfield  {title}
  {\bibinfo {title} {Universal {{Bound}} on {{Energy Cost}} of {{Bit Reset}} in
  {{Finite Time}}},\ }\href {https://doi.org/10.1103/PhysRevLett.127.190602}
  {\bibfield  {journal} {\bibinfo  {journal} {Phys. Rev. Lett.}\ }\textbf
  {\bibinfo {volume} {127}},\ \bibinfo {pages} {190602} (\bibinfo {year}
  {2021})}\BibitemShut {NoStop}%
\bibitem [{\citenamefont {Alajaji}\ \emph {et~al.}(2018)\citenamefont
  {Alajaji}, \citenamefont {Chen} \emph {et~al.}}]{alajaji2018}%
  \BibitemOpen
  \bibfield  {author} {\bibinfo {author} {\bibfnamefont {F.}~\bibnamefont
  {Alajaji}}, \bibinfo {author} {\bibfnamefont {P.-N.}\ \bibnamefont {Chen}},
  \emph {et~al.},\ }\href@noop {} {\emph {\bibinfo {title} {An Introduction to
  Single-User Information Theory}}}\ (\bibinfo  {publisher} {Springer},\
  \bibinfo {year} {2018})\BibitemShut {NoStop}%
\bibitem [{\citenamefont {Van~Vu}\ and\ \citenamefont
  {Hasegawa}(2021{\natexlab{a}})}]{vanvu2021}%
  \BibitemOpen
  \bibfield  {author} {\bibinfo {author} {\bibfnamefont {T.}~\bibnamefont
  {Van~Vu}}\ and\ \bibinfo {author} {\bibfnamefont {Y.}~\bibnamefont
  {Hasegawa}},\ }\bibfield  {title} {\bibinfo {title} {Geometrical {{Bounds}}
  of the {{Irreversibility}} in {{Markovian Systems}}},\ }\href
  {https://doi.org/10.1103/PhysRevLett.126.010601} {\bibfield  {journal}
  {\bibinfo  {journal} {Phys. Rev. Lett.}\ }\textbf {\bibinfo {volume} {126}},\
  \bibinfo {pages} {010601} (\bibinfo {year} {2021}{\natexlab{a}})}\BibitemShut
  {NoStop}%
\bibitem [{\citenamefont {Van~Vu}\ and\ \citenamefont
  {Hasegawa}(2021{\natexlab{b}})}]{vanvu2021a}%
  \BibitemOpen
  \bibfield  {author} {\bibinfo {author} {\bibfnamefont {T.}~\bibnamefont
  {Van~Vu}}\ and\ \bibinfo {author} {\bibfnamefont {Y.}~\bibnamefont
  {Hasegawa}},\ }\bibfield  {title} {\bibinfo {title} {Lower {{Bound}} on
  {{Irreversibility}} in {{Thermal Relaxation}} of {{Open Quantum Systems}}},\
  }\href {https://doi.org/10.1103/PhysRevLett.127.190601} {\bibfield  {journal}
  {\bibinfo  {journal} {Phys. Rev. Lett.}\ }\textbf {\bibinfo {volume} {127}},\
  \bibinfo {pages} {190601} (\bibinfo {year} {2021}{\natexlab{b}})}\BibitemShut
  {NoStop}%
\bibitem [{\citenamefont {Van~Vu}\ and\ \citenamefont
  {Saito}(2022)}]{vanvu2022}%
  \BibitemOpen
  \bibfield  {author} {\bibinfo {author} {\bibfnamefont {T.}~\bibnamefont
  {Van~Vu}}\ and\ \bibinfo {author} {\bibfnamefont {K.}~\bibnamefont {Saito}},\
  }\bibfield  {title} {\bibinfo {title} {Finite-{{Time Quantum Landauer
  Principle}} and {{Quantum Coherence}}},\ }\href
  {https://doi.org/10.1103/PhysRevLett.128.010602} {\bibfield  {journal}
  {\bibinfo  {journal} {Phys. Rev. Lett.}\ }\textbf {\bibinfo {volume} {128}},\
  \bibinfo {pages} {010602} (\bibinfo {year} {2022})}\BibitemShut {NoStop}%
\bibitem [{\citenamefont {Nicholson}\ \emph {et~al.}(2020)\citenamefont
  {Nicholson}, \citenamefont {{Garc{\'i}a-Pintos}}, \citenamefont {{del
  Campo}},\ and\ \citenamefont {Green}}]{nicholson2020}%
  \BibitemOpen
  \bibfield  {author} {\bibinfo {author} {\bibfnamefont {S.~B.}\ \bibnamefont
  {Nicholson}}, \bibinfo {author} {\bibfnamefont {L.~P.}\ \bibnamefont
  {{Garc{\'i}a-Pintos}}}, \bibinfo {author} {\bibfnamefont {A.}~\bibnamefont
  {{del Campo}}},\ and\ \bibinfo {author} {\bibfnamefont {J.~R.}\ \bibnamefont
  {Green}},\ }\bibfield  {title} {\bibinfo {title} {Time\textendash information
  uncertainty relations in thermodynamics},\ }\href
  {https://doi.org/10.1038/s41567-020-0981-y} {\bibfield  {journal} {\bibinfo
  {journal} {Nat. Phys.}\ }\textbf {\bibinfo {volume} {16}},\ \bibinfo {pages}
  {1211} (\bibinfo {year} {2020})}\BibitemShut {NoStop}%
\bibitem [{\citenamefont {{Garc{\'i}a-Pintos}}\ \emph
  {et~al.}(2022)\citenamefont {{Garc{\'i}a-Pintos}}, \citenamefont {Nicholson},
  \citenamefont {Green}, \citenamefont {{del Campo}},\ and\ \citenamefont
  {Gorshkov}}]{garcia-pintos2022}%
  \BibitemOpen
  \bibfield  {author} {\bibinfo {author} {\bibfnamefont {L.~P.}\ \bibnamefont
  {{Garc{\'i}a-Pintos}}}, \bibinfo {author} {\bibfnamefont {S.~B.}\
  \bibnamefont {Nicholson}}, \bibinfo {author} {\bibfnamefont {J.~R.}\
  \bibnamefont {Green}}, \bibinfo {author} {\bibfnamefont {A.}~\bibnamefont
  {{del Campo}}},\ and\ \bibinfo {author} {\bibfnamefont {A.~V.}\ \bibnamefont
  {Gorshkov}},\ }\bibfield  {title} {\bibinfo {title} {Unifying {{Quantum}} and
  {{Classical Speed Limits}} on {{Observables}}},\ }\href
  {https://doi.org/10.1103/PhysRevX.12.011038} {\bibfield  {journal} {\bibinfo
  {journal} {Phys. Rev. X}\ }\textbf {\bibinfo {volume} {12}},\ \bibinfo
  {pages} {011038} (\bibinfo {year} {2022})}\BibitemShut {NoStop}%
\bibitem [{\citenamefont {Falasco}\ and\ \citenamefont
  {Esposito}(2020)}]{falasco2020}%
  \BibitemOpen
  \bibfield  {author} {\bibinfo {author} {\bibfnamefont {G.}~\bibnamefont
  {Falasco}}\ and\ \bibinfo {author} {\bibfnamefont {M.}~\bibnamefont
  {Esposito}},\ }\bibfield  {title} {\bibinfo {title} {Dissipation-{{Time
  Uncertainty Relation}}},\ }\href
  {https://doi.org/10.1103/PhysRevLett.125.120604} {\bibfield  {journal}
  {\bibinfo  {journal} {Phys. Rev. Lett.}\ }\textbf {\bibinfo {volume} {125}},\
  \bibinfo {pages} {120604} (\bibinfo {year} {2020})}\BibitemShut {NoStop}%
\bibitem [{\citenamefont {Dechant}(2022)}]{dechant2022a}%
  \BibitemOpen
  \bibfield  {author} {\bibinfo {author} {\bibfnamefont {A.}~\bibnamefont
  {Dechant}},\ }\bibfield  {title} {\bibinfo {title} {Minimum entropy
  production, detailed balance and {{Wasserstein}} distance for continuous-time
  {{Markov}} processes},\ }\href {https://doi.org/10.1088/1751-8121/ac4ac0}
  {\bibfield  {journal} {\bibinfo  {journal} {J. Phys. A: Math. Theor.}\
  }\textbf {\bibinfo {volume} {55}},\ \bibinfo {pages} {094001} (\bibinfo
  {year} {2022})}\BibitemShut {NoStop}%
\bibitem [{\citenamefont {{del Campo}}\ \emph {et~al.}(2013)\citenamefont {{del
  Campo}}, \citenamefont {Egusquiza}, \citenamefont {Plenio},\ and\
  \citenamefont {Huelga}}]{delcampo2013}%
  \BibitemOpen
  \bibfield  {author} {\bibinfo {author} {\bibfnamefont {A.}~\bibnamefont {{del
  Campo}}}, \bibinfo {author} {\bibfnamefont {I.~L.}\ \bibnamefont
  {Egusquiza}}, \bibinfo {author} {\bibfnamefont {M.~B.}\ \bibnamefont
  {Plenio}},\ and\ \bibinfo {author} {\bibfnamefont {S.~F.}\ \bibnamefont
  {Huelga}},\ }\bibfield  {title} {\bibinfo {title} {Quantum {{Speed Limits}}
  in {{Open System Dynamics}}},\ }\href
  {https://doi.org/10.1103/PhysRevLett.110.050403} {\bibfield  {journal}
  {\bibinfo  {journal} {Phys. Rev. Lett.}\ }\textbf {\bibinfo {volume} {110}},\
  \bibinfo {pages} {050403} (\bibinfo {year} {2013})}\BibitemShut {NoStop}%
\end{thebibliography}

%

\vspace{5em}
\newpage
\widetext
\appendix
\section{\Large Supplemental Materials: Speed limit,  dissipation bound and dissipation-time trade-off in thermal relaxation processes}

\setcounter{equation}{0}
\setcounter{figure}{0}
\setcounter{table}{0}
\setcounter{page}{1}
\makeatletter
\renewcommand{\theequation}{S\arabic{equation}}
\renewcommand{\thefigure}{S\arabic{figure}}
\renewcommand{\bibnumfmt}[1]{[S#1]}
\renewcommand{\citenumfont}[1]{S#1}

\section{Derivation of Eq. (8)}
Consider a two-state systems, where for simplicity we define $k_{1}=W_{21}$ and $k_{2}=W_{12}$.
 The transition rate matrix can then be represented as:
 \begin{equation}
 \boldsymbol{W} = \begin{bmatrix}
 -k_{1} & k_{2} \\ k_{1} & -k_{2}
 \end{bmatrix}.
 \end{equation}
The two eigenstates $\lambda_n$ and right-eigenvectors $\boldsymbol{\eta}_n$ ($n=1,2$) are, respectively,
\begin{equation}
\lambda_{1}=0, \quad \lambda_{2} =-\mathcal{W}, \quad \mathcal{W}=\left(k_{1}+k_{2}\right)
\end{equation}
and
\begin{equation}
\boldsymbol{\eta}_{1}=\left[\begin{array}{c}p_1^\text{(ss)} \\ 1-p_1^\text{(ss)}\end{array}\right], \quad \boldsymbol{\eta}_2 = \left[\begin{array}{c}1 \\ -1\end{array}\right].
\end{equation}

By utilizing the eigenvalue method for solving systems of ordinary differential equations, we obtain
\begin{equation}
\left \{
\begin{aligned}
&\boldsymbol{p}^\text{(i)} = c_1 \boldsymbol{\eta}_1 + c_2 \boldsymbol{\eta}_2 \\
&\boldsymbol{p}^\text{(f)} = c_1 \boldsymbol{\eta}_1 + c_2 e^{\lambda_2 \tau} \boldsymbol{\eta}_2 
\end{aligned}
\right .
\end{equation}

Given the two endpoints,  solving the linear equations gives the steady states expressed in terms of $\mathcal{W} \tau$,
\begin{equation}
   p_1^\text{(ss)} = p_1^\text{(i)} + \frac{p_1^\text{(f)}-p_1^\text{(i)}}{1-e^{-\mathcal{W} \tau}} ,
\end{equation}
with $p_2^\text{(ss)}=1-p_1^\text{(ss)}$.

\section{Code for verifying Eqs. (11) and (13)}
\begin{python}
import numpy as np
from scipy.linalg import expm, eig
from scipy import special, stats
import matplotlib.pyplot as plt
plt.rcParams['font.family'] = 'Times New Roman'
plt.rcParams['mathtext.fontset'] = 'cm'
plt.rc('text', usetex=True)

def compute_steady_state_distribution(transition_matrix):
    """
    Computes the steady state distribution.

    Args:
        transition_matrix (numpy.ndarray): The transition matrix of the Markov chain.

    Returns:
        numpy.ndarray: The steady state distribution of the Markov chain.
    """
    eigen_values, eigen_vectors = eig(transition_matrix)
    idx = eigen_values.argsort()[::-1]
    eigen_values = eigen_values[idx]
    eigen_vectors = eigen_vectors[:, idx]
    steady_state_distribution = eigen_vectors[:, 0].real
    steady_state_distribution = steady_state_distribution / steady_state_distribution.sum()
    return steady_state_distribution

def compute_entropy_production(dimension):
    """
    Computes the non-adiabatic entropy production and bounds.

    Args:
        dimension (int): number of states (dimension of transition rate matrix).

    Returns:
        list: A list containing two entropy productions.
    """
    initial_distribution = np.random.rand(dimension)
    initial_distribution = initial_distribution / initial_distribution.sum()
    tau = 1.0 * np.random.rand()
    transition_matrix = np.zeros((dimension, dimension))
    for j in range(dimension):
        for i in range(dimension):
            if i != j:
                transition_matrix[i][j] = np.random.rand()
        transition_matrix[j][j] = -transition_matrix[:, j].sum()
    final_distribution = np.matmul(expm(transition_matrix * tau), initial_distribution)
    mid_distribution = np.matmul(expm(transition_matrix * tau/2.0), initial_distribution)
    steady_state_distribution = compute_steady_state_distribution(transition_matrix)
    entropy_production = stats.entropy(initial_distribution, steady_state_distribution) - stats.entropy(final_distribution, steady_state_distribution)
    entropy_production_mid = stats.entropy(initial_distribution, steady_state_distribution) - stats.entropy(mid_distribution, steady_state_distribution)
    KL = stats.entropy(initial_distribution, final_distribution)
#     pseudo-coarse-grained (PCG) dyamics
    num = int(1 + (dimension - 1) * np.random.rand())
    PCG_initial_distribution = initial_distribution[:num].sum()
    PCG_initial_distribution = np.array([PCG_initial_distribution, 1 - PCG_initial_distribution])
    PCG_final_distribution = final_distribution[:num].sum()
    PCG_final_distribution = np.array([PCG_final_distribution, 1 - PCG_final_distribution])
    PCG_steady_state_distribution_1 = PCG_initial_distribution[0] + (PCG_final_distribution[0] - PCG_initial_distribution[0]) / (1 - np.exp(transition_matrix.trace() * tau))
    PCG_steady_state_distribution_2 = 1.0 - PCG_steady_state_distribution_1
    PCG_steady_state_distribution = np.array([PCG_steady_state_distribution_1, PCG_steady_state_distribution_2])
    entropy_production_PCG = stats.entropy(PCG_initial_distribution, PCG_steady_state_distribution) - stats.entropy(PCG_final_distribution, PCG_steady_state_distribution)
    return [entropy_production_PCG, entropy_production, entropy_production/KL, entropy_production_mid/KL]

def plot_fig1ab(dimension=3,npoints=100):
    """
    plot Fig.1(a) and (b)

    Args:
        dimension (int): The dimension of the Markov chain.
        npoints (int): Number of data points.
    """
    # Generate the data
    data = np.zeros((npoints, 4))
    for i in range(npoints):
        data[i, :] = compute_entropy_production(dimension)

    # Plot Fig. 1(a)
    fig = plt.figure(num=1, figsize=(8, 6), dpi=300)
    ax = fig.add_subplot(111)
    ax.set_xlim([0, npoints])
    ax.set_ylim([0.5, 10])
    plt.yscale("log")
    ax.scatter(np.arange(npoints), data[:,2], s=1.5, color="black", edgecolors='none')
    ax.scatter(np.arange(npoints), data[:,3], s=0.05, color="white", edgecolors='gray')
    ax.plot(np.arange(npoints), np.ones(npoints), lw=0.5, color='red')
    plt.tick_params(labelsize=17, which="both", direction="in")
    plt.xticks([npoints])
    plt.xlabel('iteration', size=20)
    plt.ylabel(r'$\Sigma_{\mathrm{na}}/D_1$', size=20)
    plt.savefig("fig1a.png")
    plt.close()

    # Plot Fig. 1(b)
    fig = plt.figure(num=1, figsize=(8, 6), dpi=300)
    ax = fig.add_subplot(111)
    ax.set_xlim([1e-4, 1.4])
    ax.set_ylim([1e-4, 1.4])
    plt.xscale("log")
    plt.yscale("log")
    ax.scatter(data[:, 0], data[:, 1], s=0.5, color="black",edgecolors='none')
    x = np.linspace(0.0, 10, 10000)
    ax.plot(x, x, lw=0.5, color="red")
    plt.tick_params(labelsize=17, which="both", direction="in")
    plt.xlabel(r'$\Sigma_{\Omega_i}$', size=20)
    plt.ylabel(r'$\Sigma_{\mathrm{na}}$', size=20)
    plt.savefig("fig1b.png")
    # plt.show()

# Set the dimension of the Markov chain and the number of data points to generate
dimension = 3
npoints = 100000
plot_fig1ab(dimension,npoints)
\end{python}
\end{document}